\documentclass[showpacs,preprintnumbers,twocolumn,superscriptaddress,aps,nofootinbib]{revtex4-1}
\usepackage{amssymb}
\usepackage[centertags]{amsmath}
\usepackage{txfonts}
\usepackage{epsfig}
\usepackage{bm}
\usepackage{color}
\usepackage{graphicx,graphics}
\usepackage{multirow}
\usepackage{float}
\usepackage{ulem}
\usepackage{appendix}
\usepackage{longtable} 
\usepackage{xcolor} 
\usepackage{natbib}  
\usepackage{threeparttablex}
\usepackage[pdfstartview=FitH]{hyperref}
\hypersetup{colorlinks=true, citecolor=blue,linkcolor=blue,filecolor=black,urlcolor=blue}  
\allowdisplaybreaks[2]

\begin{document}

\title{Production correlation of light (hyper-)nuclei in Au-Au collisions from the RHIC Beam Energy Scan}

\author{Jiang-He Qiao}
\affiliation{School of Physics and Physical Engineering, Qufu Normal University, Shandong 273165, China}

\author{Jian-Yu Liu}
\affiliation{School of Physics and Physical Engineering, Qufu Normal University, Shandong 273165, China}

\author{Yan-Ting Feng}
\affiliation{School of Physics and Physical Engineering, Qufu Normal University, Shandong 273165, China}



\author{Feng-Lan Shao}
\affiliation{School of Physics and Physical Engineering, Qufu Normal University, Shandong 273165, China}

\author{Rui-Qin Wang}
\email {wangrq@qfnu.edu.cn}
\affiliation{School of Physics and Physical Engineering, Qufu Normal University, Shandong 273165, China}

\begin{abstract}
Based on nucleons ($p$, $n$) and hyperons ($\Lambda$, $\Omega^-$) formed at kinetic freeze-out from a quark combination model, we systematically study the production of light nuclei and hyper-nuclei in the hadronic coalescence picture.
We present the analytical formula of the nucleus momentum distribution of two-body coalescence and that of three-body coalescence.
We explain the experimental data of the transverse momentum spectra of deuteron ($d$), triton ($t$), helium-3 ($^3$He) and hypertriton ($^3_{\Lambda}$H) measured in Au-Au collisions from the RHIC Beam Energy Scan I and II,
and also provide the corresponding predictions of different $\Omega-$hypernuclei $H(p\Omega^-)$, $H(n\Omega^-)$ and $H(pn\Omega^-)$.
We further study the production correlations of different species of light (hyper-)nuclei and discuss their interesting behaviors as a function of collision energy.
\end{abstract}

\maketitle

\section{Introduction}

Light nuclei and hyper-nuclei are a special group of observables in relativistic heavy-ion collisions~\cite{Chen:2018tnh,Nemura:1999qp,Oliinychenko:2020ply}.
Due to their small binding energies, several MeV or even smaller several hundred keV~\cite{Juric:1973zq,STAR:2019wjm,ALICE:2022sco}, light (hyper-)nuclei are expected to be formed at the late stage of the evolution of a heavy-ion collision. 
So they can serve as direct probes for the freeze-out properties of the bulk strong-interaction system~\cite{Nagle:1994hm,Wang:2012eof,Bazak:2020wjn,Zhang:2023ikt}.
Meanwhile, the production of such composite clusters itself offers a natural way to constrain the information of both nucleon-hyperon interaction and possible three-body forces~\cite{NPLQCD:2012mex,Ma:2023,Wirth:2014apa,ALICE:2022boj,Haidenbauer:2025zrr}. 

Theoretical study of the production of light (hyper-)nuclei has always absorbed a lot of attention in heavy-ion collisions~\cite{Schwarzschild:1963zz,Siemens:1979dz,Aichelin:1991xy,Mattiello:1995xg,Oh:2009gx,Andronic:2017pug,Bzdak:2019pkr}.
Thermal production mechanism~\cite{Mekjian:1977ei,Andronic:2010qu,Vovchenko:2018fiy} and coalescence mechanism~\cite{Sato:1981ez,Scheibl:1998tk,Chen:2003qj} are popularly utilized to deal with the formation of such composite objects.
Transport scenario is also employed to study how light (hyper-)nuclei interact and survive in the hadronic system evolution~\cite{Danielewicz:1991dh,Oliinychenko:2018ugs,Kireyeu:2022qmv,Coci:2023daq}.
In the coalescence mechanism, the production of $d$, $t$, $^3$He, and $^3_{\Lambda}$H, closely related with their internal structures, has been found to possess some unique characteristics.
For example, more production suppression of light nuclei in smaller collision system~\cite{Sun:2018mqq} and softer transverse momentum spectrum for $t$, $^3$He, and $^3_{\Lambda}$H with larger root-mean-square radius~\cite{Wang:2023rpd,Liu:2024ygk}.
Such characteristics are powerful signals of discriminating the production mechanism of these composite objects, which remains a subject of debate in both nuclear community and particle physics.

Production correlations of various species of light (hyper-)nuclei as well as those between nuclei and primordial nucleons/hyperons, are a class of more sensitive quantities for decoding characteristic properties of production mechanisms.
They may exhibit lush behaviors due to influences of relatively different and large sizes of various nuclei, compared to ordinary baryons and mesons. 
On the basis of overall measurements from the ALICE collaboration at the CERN Large Hadron Collider (LHC)~\cite{ALICE:2022veq,ALICE:2022sco,ALICE:2024koa,ALICE:2020mfd}, in our recent work~\cite{Wang:2024hok}, we developed a coalescence model with the isospin symmetry assumption to study the production of light nuclei, $^3_{\Lambda}$H and $\Omega-$hypernuclei in Pb-Pb collisions at $\sqrt{s_{NN}}=5.02$ TeV. 
We found two groups of characteristic correlations: the average transverse momentum ratios and centrality-dependent yield ratios of light nuclei to protons and hyper-nuclei to hyperons.
These ratios just offset the differences of the primordial $p$, $\Lambda$, and $\Omega^-$, and so were much potent to test the existence of a universal production mechanism for different nuclei~\cite{Wang:2024hok}.

Very recently, the STAR experiment at the BNL Relativistic Heavy Ion Collider (RHIC) has published the experimental data of $^3$He and $^3_{\Lambda}$H in Beam Energy Scan (BES) II~\cite{liu2025measurement,Li:2025jaz}.
The data of $d$ and $t$ have been previously collected in BES I~\cite{STAR:2019sjh}. 
Except precisely measuring these known light (hyper-)nuclei, the STAR collaboration has also executed a search for other hyper-nuclei, and their results favor the existence of $p\Omega^-$ dibaryon state~\cite{STAR:2018uho}.
Furthermore, future facilities such as the China HyperNuclear Spectrometer (CHNS) at the High Intensity heavy-ion Accelerator Facility (HIAF)~\cite{Zhou:2022pxl} and Compressed Baryonic Matter (CBM) detector at Facility for Antiproton and Ion Research (FAIR)~\cite{CBM:2016kpk} will provide more measurements in heavy-ion collisions with relatively low collision energies, about GeV magnitude and lower.
All of these facilities push the experimental study of light (hyper-)nuclei in heavy-ion collisions to usher in an era of precise measurements not only for multiform species but also in a wide collision energy range.
This encourages us to explore the energy dependence of the production correlations of various composite nuclei in light and strange sectors from the theoretical aspect.
Such study can help to deeply reveal the universal production mechanism of various light (hyper-)nuclei, and also can decode influences of different bulk hadronic systems on nucleus production.

In this article, considering the isospin asymmetry of protons and neutrons resulted from the colliding gold nuclei and correcting those nucleons and hyperons from weak and electromagnetic decays with a Quark Combination Model (QCM)~\cite{Feng:2022dln}, we extend the coalescence model developed for heavy-ion collisions at the LHC~\cite{Wang:2023rpd,Wang:2024hok} to Au-Au collisions at the RHIC.
We study the transverse momentum ($p_T$) spectra, the yield rapidity densities at midrapidity ($dN/dy$), and the average transverse momenta ($\langle p_T\rangle$) of $d$, $t$, $^3$He, $^3_{\Lambda}$H and $\Omega-$hypernuclei $H(p\Omega^-)$, $H(n\Omega^-)$, $H(pn\Omega^-)$ in central Au-Au collisions at $\sqrt{s_{NN}}=7.7,~9.2,~11.5,~14.5,~17.3,~19.6,~27,~39,~54.4,~62.4$, and 200 GeV.
We particularly study the production correlations of different species of light (hyper-)nuclei and present their behaviors as a function of collision energy.
We find the location of two-nucleus yield ratio is powerful to exposure the nucleus relative size.
We also find the average transverse momenta of both light nuclei and hyper-nuclei, except the $^3_{\Lambda}$H, obey the mass order from $\sqrt{s_{NN}}=7.7$ GeV to 200 GeV.

The rest of the paper is organized as follows. 
In Sec.~\ref{model}, we introduce the coalescence model.
In Sec.~\ref{pTresults}, we study the $p_T$ spectra, $dN/dy$ and $\langle p_T\rangle$ of $d$, $t$, $^3$He, $^3_{\Lambda}$H and $\Omega-$hypernuclei $H(p\Omega^-)$, $H(n\Omega^-)$, $H(pn\Omega^-)$ in central Au-Au collisions at $\sqrt{s_{NN}}=7.7,~9.2,~11.5,~14.5,~17.3,~19.6,~27,~39,~54.4,~62.4$, and 200 GeV.
We specially present the influences of different internal structures of $^3_{\Lambda}$H on its productivity at different collision energies.
In Sec.~\ref{Cor-results}, we study two groups of global correlations and analyze their collision energy dependent behaviors.
One group are yield ratios of various light (hyper-)nuclei and the other are average transverse momentum relations.
In Sec.~\ref{summary}, we summarize our work.

\section{The coalescence model}   \label{model}

We consider the isospin asymmetry to extend the analytical coalescence model developed at the LHC in Refs.~\cite{Wang:2023rpd,Wang:2024hok}, to manage the production of different light nuclei and hyper-nuclei at the RHIC.
The current coalescence model absorbs the finite emission duration in an effective volume instead of carrying out the time evolution step by step.
This can make the analytical and intuitive insights possible and provide an effective method to extract space scale information of the hadronic system formed at the light (hyper-)nucleus freeze-out.
We first present the formalism of two bodies coalescing into nuclei like $d$ and nucleon$-\Omega$ dibaryon state. 
Then we give an analytical expression of three bodies coalescing into $t$, $^3$He, and their partners with strangeness quantum number.

\subsection{Formalism of two-body coalescence}      \label{2hcoHj}

We begin with a hadronic system produced at the final stage of the evolution of high energy collision, and suppose the light (hyper-)nucleus $H_{j}$ is formed via the coalescence of two hadronic bodies $h_1$ and $h_2$.
We use $f_{H_j}(\bm{p})$ to denote the three-dimensional momentum distribution of the produced $H_{j}$ and it is given by
{\setlength\arraycolsep{0pt}
\begin{eqnarray}
f_{H_{j}}(\bm{p}) =&&  \int d\bm{x}_1d\bm{x}_2 d\bm{p}_1 d\bm{p}_2  f_{h_1h_2}(\bm{x}_1,\bm{x}_2;\bm{p}_1,\bm{p}_2)   \nonumber  \\
&&\times  \mathcal {R}_{H_{j}}(\bm{x}_1,\bm{x}_2;\bm{p}_1,\bm{p}_2,\bm{p}).     \label{eq:fHj2hgeneral} 
\end{eqnarray} }%
$f_{h_1h_2}(\bm{x}_1,\bm{x}_2;\bm{p}_1,\bm{p}_2)$ is the two-hadron joint coordinate-momentum distribution and
$\mathcal {R}_{H_{j}}(\bm{x}_1,\bm{x}_2;\bm{p}_1,\bm{p}_2,\bm{p})$ is the kernel function of the $H_{j}$.
Here and from now on we use bold symbols to denote three-dimensional coordinate or momentum vectors.

The two-hadron joint distribution $f_{h_1h_2}(\bm{x}_1,\bm{x}_2;\bm{p}_1,\bm{p}_2)$ is the number density that satisfies
{\setlength\arraycolsep{0pt}
\begin{eqnarray}
 \int d\bm{x}_1d\bm{x}_2 d\bm{p}_1 d\bm{p}_2  f_{h_1h_2}(\bm{x}_1,\bm{x}_2;\bm{p}_1,\bm{p}_2)  = N_{h_1h_2},    
\end{eqnarray} }%
where $N_{h_1h_2}$ is the number of all possible $h_1h_2$-pairs and it is equal to $N_{h_1}N_{h_2}$.
$N_{h_1}$ and $N_{h_2}$ are the number of $h_1$ hadrons and that of $h_2$ hadrons in the considered hadronic system.
For the convenience of comparison, we rewrite
{\setlength\arraycolsep{0pt}
\begin{eqnarray}
 f_{h_1h_2}(\bm{x}_1,\bm{x}_2;\bm{p}_1,\bm{p}_2)  = N_{h_1h_2}  f^{(n)}_{h_1h_2}(\bm{x}_1,\bm{x}_2;\bm{p}_1,\bm{p}_2),    
\end{eqnarray} }%
so that the joint distribution is normalized to unity where we denote by the superscript `$(n)$'.
In terms of the normalized joint coordinate-momentum distribution, we have
{\setlength\arraycolsep{0pt}
\begin{eqnarray}
f_{H_{j}}(\bm{p})=&& N_{h_1h_2} \int d\bm{x}_1d\bm{x}_2 d\bm{p}_1 d\bm{p}_2  f^{(n)}_{h_1h_2}(\bm{x}_1,\bm{x}_2;\bm{p}_1,\bm{p}_2)    \nonumber  \\
&&\times   \mathcal {R}_{H_{j}}(\bm{x}_1,\bm{x}_2;\bm{p}_1,\bm{p}_2,\bm{p}).      \label{eq:fHj2hgeneral1}               
\end{eqnarray} }%

The kernel function $\mathcal {R}_{H_{j}}(\bm{x}_1,\bm{x}_2;\bm{p}_1,\bm{p}_2,\bm{p})$ denotes the probability density for $h_1$, $h_2$ with
momenta $\bm{p}_1$ and $\bm{p}_2$ at $\bm{x}_1$ and $\bm{x}_2$ to combine into an $H_{j}$ of momentum $\bm{p}$.
Its precise expression should be constrained by such as the momentum conservation, and constraints due to intrinsic quantum numbers e.g., spin~\cite{Wang:2020zaw,Zhao:2022xkz,Wang:2023rpd,Wang:2024hok}.
To take these constraints into account explicitly, we rewrite the kernel function as
{\setlength\arraycolsep{0pt}
\begin{eqnarray}
\mathcal {R}_{H_{j}}(\bm{x}_1,\bm{x}_2;\bm{p}_1,\bm{p}_2,\bm{p}) =&& g_{H_{j}} \mathcal {R}_{H_{j}}^{(x,p)}(\bm{x}_1,\bm{x}_2;\bm{p}_1,\bm{p}_2)     \nonumber  \\
&& \times  \delta(\displaystyle{\sum^2_{i=1}} \bm{p}_i-\bm{p}).   \label{eq:RHj2h}  
\end{eqnarray} }%
The spin degeneracy factor $g_{H_{j}} = (2J_{H_{j}}+1) /[\prod \limits_{i=1}^2(2J_{h_i}+1)]$, where $J_{H_{j}}$ is the spin of $H_{j}$ and $J_{h_i}$ is that of the primordial $h_i$. 
The Dirac $\delta$ function guarantees the momentum conservation in the coalescence process.
The remaining $\mathcal {R}_{H_{j}}^{(x,p)}(\bm{x}_1,\bm{x}_2;\bm{p}_1,\bm{p}_2)$ can be solved from the Wigner transformation as the $H_{j}$ wave function is given.
Considering the wave function of a spherical harmonic oscillator is particularly tractable and useful for analytical insight, we adopt this profile as in Refs.~\cite{Chen:2003ava,Ko:2012lhi,Zhu:2015voa} and have
{\setlength\arraycolsep{0pt}
\begin{eqnarray}
 \mathcal {R}^{(x,p)}_{H_{j}}(\bm{x}_1,\bm{x}_2;\bm{p}_1,\bm{p}_2) = 8e^{-\frac{(\bm{x}'_1-\bm{x}'_2)^2}{2\sigma^2}}
	e^{-\frac{2\sigma^2(m_2\bm{p}'_{1}-m_1\bm{p}'_{2})^2}{(m_1+m_2)^2}}. ~~~~     \label{eq:RHj2hxp} 
\end{eqnarray} }%
The superscript `$'$' in the coordinate or momentum variables denotes the hadron coordinate or momentum in the rest frame of $h_1h_2$-pair.
$m_1$ and $m_2$ are the mass of $h_1$ and that of $h_2$.
The width parameter $\sigma=\sqrt{\frac{2(m_1+m_2)^2}{3(m_1^2+m_2^2)}} R_{H_{j}}$, and $R_{H_{j}}$ is the root-mean-square radius of $H_{j}$.

Substituting Eqs.~(\ref{eq:RHj2h}) and (\ref{eq:RHj2hxp}) into Eq.~(\ref{eq:fHj2hgeneral1}), we have
{\setlength\arraycolsep{0.2pt}
\begin{eqnarray}
 f_{H_{j}}(\bm{p})=&& N_{h_1h_2} g_{H_{j}} \int d\bm{x}_1d\bm{x}_2d\bm{p}_1d\bm{p}_2 f^{(n)}_{h_1h_2}(\bm{x}_1,\bm{x}_2;\bm{p}_1,\bm{p}_2)    \nonumber  \\
&& \times
8e^{-\frac{(\bm{x}'_1-\bm{x}'_2)^2}{2\sigma^2}}   e^{-\frac{2\sigma^2(m_2\bm{p}'_{1}-m_1\bm{p}'_{2})^2}{(m_1+m_2)^2}}
	\delta(\displaystyle{\sum^2_{i=1}} \bm{p}_i-\bm{p}).    \label{eq:fHj2h}  
\end{eqnarray} }%
This is the general formalism of $H_j$ production via the coalescence of two hadronic bodies $h_1$ and $h_2$.

Noticing that for light (hyper-)nuclei the root-mean-square radius $R_{H_{j}}$ is always about or larger than 2 fm~\cite{Angeli:2013epw}, $\sigma$ is even larger than $R_{H_{j}}$.
So the Gaussian width in the momentum-dependent part of the kernel function in Eq.~(\ref{eq:fHj2h}) has a small value, about or smaller than 0.1 GeV/c.
Therefore, we approximate the Gaussian form of the momentum-dependent kernel function to be a $\delta$ function form as follows
{\setlength\arraycolsep{0.2pt}
\begin{equation}
e^{-\frac{(\bm{p}'_{1}-\frac{m_1}{m_2}\bm{p}'_{2})^2} {(1+\frac{m_1}{m_2})^2/ (2\sigma^2)}} \approx
\left[ \frac{\sqrt{\pi}}{\sqrt{2}\sigma} (1+\frac{m_1}{m_2})  \right]^3 \delta(\bm{p}'_{1}-\frac{m_1}{m_2}\bm{p}'_{2}).    \label{eq:deltaapp}  
\end{equation} }%
The robustness of this $\delta$ function approximation has been checked at the outset of the analytical coalescence model in our previous work~\cite{Wang:2020zaw}.
Substituting Eq.~(\ref{eq:deltaapp}) into Eq.~(\ref{eq:fHj2h}) and integrating $\bm{p}_1$ and $\bm{p}_2$, we can obtain
{\setlength\arraycolsep{0.2pt}
\begin{eqnarray}
 f_{H_{j}}(\bm{p})=&& N_{h_1h_2} g_{H_{j}} \int d\bm{x}_1d\bm{x}_2d\bm{p}_1d\bm{p}_2  f^{(n)}_{h_1h_2}(\bm{x}_1,\bm{x}_2;\bm{p}_1,\bm{p}_2)    \nonumber  \\
&&\times  8 e^{-\frac{(\bm{x}'_1-\bm{x}'_2)^2}{2 \sigma^2}}   (\frac{\sqrt{\pi}}{\sqrt{2} \sigma})^3 (1+\frac{m_1}{m_2})^3  \delta(\bm{p}'_{1}-\frac{m_1}{m_2} \bm{p}'_{2})  \nonumber  \\
&& \times  \delta(\displaystyle{\sum^2_{i=1}} \bm{p}_i-\bm{p})                 \nonumber    \\                             
=&&  N_{h_1h_2} g_{H_{j}} \int d\bm{x}_1d\bm{x}_2d\bm{p}_1d\bm{p}_2 f^{(n)}_{h_1h_2}(\bm{x}_1,\bm{x}_2;\bm{p}_1,\bm{p}_2)     \nonumber  \\
&& \times  8e^{-\frac{(\bm{x}'_1-\bm{x}'_2)^2}{2 \sigma^2}}  
(\frac{\sqrt{\pi}}{\sqrt{2} \sigma})^3 (1+\frac{m_1}{m_2})^3
\gamma \delta(\bm{p}_{1}-\frac{m_1}{m_2} \bm{p}_{2})   \nonumber  \\
&&  \times \delta(\displaystyle{\sum^2_{i=1}} \bm{p}_i-\bm{p})         \nonumber   \\              
=&&N_{h_1h_2}  g_{H_{j}} \gamma (\frac{\sqrt{\pi}}{\sqrt{2} \sigma})^3  \times 8  \int d\bm{x}_1d\bm{x}_2  e^{-\frac{(\bm{x}'_1-\bm{x}'_2)^2}{2 \sigma^2}}     \nonumber  \\
&& \times   f^{(n)}_{h_1h_2}(\bm{x}_1,\bm{x}_2;\frac{m_1 \bm{p}}{m_1+m_2},\frac{m_2 \bm{p}}{m_1+m_2}) .    \label{eq:fH-p}   
\end{eqnarray} }%
The $\gamma$ is the Lorentz contraction factor corresponding to the three-dimensional velocity $\bm{\beta}$ of the center-of-mass frame of $h_1h_2$-pair in the laboratory frame.
Here the momentum transformation parallel to $\bm{\beta}$ is $p'_{1//}-\frac{m_1}{m_2} p'_{2//}=\frac{1}{\gamma} (p_{1//}-\frac{m_1}{m_2} p_{2//})$ 
and that perpendicular to $\bm{\beta}$ is invariant. 

Changing coordinate variables in Eq.~(\ref{eq:fH-p}) to be $\bm{X}=\frac{\sqrt{2}(m_1\bm{x}_1+m_2\bm{x}_2)}{m_1+m_2}$ and $\bm{r}=\frac{\bm{x}_1-\bm{x}_2} {\sqrt{2}}$, we have  
{\setlength\arraycolsep{0pt}
\begin{eqnarray}
 f_{H_{j}}(\bm{p})=&& N_{h_1h_2} g_{H_{j}} \gamma  (\frac{\sqrt{\pi}}{\sqrt{2} \sigma})^3    \nonumber  \\
&& \times  8\int d\bm{X}d\bm{r}  f^{(n)}_{h_1h_2}(\bm{X},\bm{r};\frac{m_1 \bm{p}}{m_1+m_2},\frac{m_2 \bm{p}}{m_1+m_2})    e^{-\frac{\bm{r}'^2}{\sigma^2}}. \nonumber  \\  \label{eq:fH-Xr}  
\end{eqnarray} }%
As the strong interaction and the coalescence are local, we neglect the effect of collective motion on the center of mass coordinate and assume it is factorized, i.e.,
{\setlength\arraycolsep{0pt}
\begin{eqnarray}
&&   f^{(n)}_{h_1h_2}(\bm{X},\bm{r};\frac{m_1 \bm{p}}{m_1+m_2},\frac{m_2 \bm{p}}{m_1+m_2})    \nonumber  \\
 =&&  f^{(n)}_{h_1h_2}(\bm{X})   f^{(n)}_{h_1h_2}(\bm{r};\frac{m_1 \bm{p}}{m_1+m_2},\frac{m_2 \bm{p}}{m_1+m_2}).  \label{eq:fh1h2-Xr}  
\end{eqnarray} }%
Substituting Eq.~(\ref{eq:fh1h2-Xr}) into Eq.~(\ref{eq:fH-Xr}), we have
{\setlength\arraycolsep{0pt}
\begin{eqnarray}
 f_{H_{j}}(\bm{p}) = && N_{h_1h_2}g_{H_{j}}  \gamma  (\frac{\sqrt{\pi}}{\sqrt{2} \sigma})^3    \nonumber  \\
&& \times  8 \int d\bm{r} f^{(n)}_{h_1h_2}(\bm{r};\frac{m_1 \bm{p}}{m_1+m_2},\frac{m_2 \bm{p}}{m_1+m_2})    e^{-\frac{\bm{r}'^2}{\sigma^2}}.~~~   \label{eq:fH-r} 
\end{eqnarray} }%

We adopt the frequently-used Gaussian form for the relative coordinate distribution as in such as Refs.~\cite{Mrowczynski:2016xqm,Kisiel:2014upa,ALICE:2015tra}, i.e.,
{\setlength\arraycolsep{0pt}
\begin{eqnarray}
&& f^{(n)}_{h_1h_2}(\bm{r};\frac{m_1\bm{p}}{m_1+m_2} ,\frac{m_2 \bm{p}}{m_1+m_2}) =  \frac{1}{\left[\pi C_0 R_f^2(\bm{p})\right]^{3/2}}    \nonumber  \\
&&~~~~~~~~ \times
  e^{-\frac{\bm{r}^2}{C_0 R_f^2(\bm{p})}}   f^{(n)}_{h_1h_2}(\frac{m_1\bm{p}}{m_1+m_2} ,\frac{m_2 \bm{p}}{m_1+m_2}).     \label{eq:fh1h2-rp}    
\end{eqnarray} }%
Here, $R_f(\bm{p})$ is the effective radius of the hadronic source system at the $H_j$ freeze-out.
$C_0$ is introduced to make $\bm{r}^2/C_0$ to be the square of one-half of the relative position and it is 2~\cite{Mrowczynski:2016xqm,Kisiel:2014upa,ALICE:2015tra}.
In this way $R_f(\bm{p})$ is just the Hanbury-Brown-Twiss interferometry radius, which can also be extracted from the two-particle femtoscopic correlations~\cite{Kisiel:2014upa,ALICE:2015tra}.

With instantaneous coalescence in the rest frame of $h_1h_2$-pair, i.e., $\Delta t'=0$, we get the coordinate transformation
\begin{eqnarray}
\bm{r} = \bm{r}' +(\gamma-1)\frac{\bm{r}'\cdot \bm{\beta}}{\beta^2}\bm{\beta}.    \label{eq:LorentzTr}  
\end{eqnarray}%
The instantaneous coalescence is a basic assumption in coalescence-like models where the overlap of the nucleus Wigner phase-space density with the constituent phase-space distributions is adopted~\cite{Chen:2003ava}.
Considering the coalescence criterion judging in the rest frame is more general than in the laboratory frame, we choose the instantaneous coalescence in the rest frame of $h_1h_2$-pair, as in Refs.~\cite{Gutbrod:1976zzr,Chen:2003ava}.
Substituting Eq.~(\ref{eq:fh1h2-rp}) into Eq.~(\ref{eq:fH-r}) and using Eq.~(\ref{eq:LorentzTr}) to integrate from the relative coordinate variable, we obtain 
{\setlength\arraycolsep{0pt}
\begin{eqnarray}
f_{H_{j}}(\bm{p}) =&& \frac{8 \pi^{3/2} g_{H_{j}} \gamma}{2^{3/2}\left[C_0 R_f^2(\bm{p})+\sigma^2\right] \sqrt{C_0 [R_f(\bm{p})/\gamma]^2+\sigma^2}}    \nonumber  \\
&& \times  f_{h_1h_2}(\frac{m_1 \bm{p}}{m_1+m_2},\frac{m_2 \bm{p}}{m_1+m_2}).   \label{eq:fHjh1h2} 
\end{eqnarray} }%
Ignoring correlations between $h_1$ and $h_2$, we have the three-dimensional momentum distribution of the $H_j$ as
{\setlength\arraycolsep{0pt}
\begin{eqnarray}
f_{H_{j}}(\bm{p}) =&& \frac{8 \pi^{3/2} g_{H_{j}} \gamma}{2^{3/2}\left[C_0 R_f^2(\bm{p})+\sigma^2\right] \sqrt{C_0 [R_f(\bm{p})/\gamma]^2+\sigma^2}}    \nonumber  \\
&& \times  f_{h_1}(\frac{m_1\bm{p}}{m_1+m_2}) f_{h_2}(\frac{m_2\bm{p}}{m_1+m_2}).    \label{eq:fHj2h-approx}
\end{eqnarray} }%

Denoting the Lorentz invariant momentum distribution $E\dfrac{d^{3}N}{d\bm{p}^3}=\dfrac{d^{2}N}{2\pi p_{T}dp_{T}dy}$ with $f^{\text{(inv)}}$, we finally have
{\setlength\arraycolsep{0.2pt}
\begin{eqnarray}
&&f_{H_j}^{\text{(inv)}}(p_T,y) =\frac{8\pi^{3/2} g_{H_{j}} }{2^{3/2}\left[C_0 R_f^2(p_T,y)+\sigma^2\right] \sqrt{C_0 \frac{R^2_f(p_T,y)}{\gamma^2}+\sigma^2}}      \nonumber  \\
&& ~~~~~ \times \frac{m_{H_j}}{m_1m_2} f_{h_1}^{\text{(inv)}}(\frac{m_1p_{T}}{m_1+m_2},y)  f_{h_2}^{\text{(inv)}}(\frac{m_2p_{T}}{m_1+m_2},y),     \label{eq:pt-Hj2h}
\end{eqnarray} }%
where $m_{H_j}$ is the $H_j$ mass and $y$ is longitudinal rapidity.

\begin{widetext}
\subsection{Formalism of three-body coalescence}

For light (hyper-)nuclei $H_{j}$ formed via the coalescence of three bodies $h_1$, $h_2$ and $h_3$, the production formalism is very similar to that of two-body coalescence and is simply given in the following.
The momentum distribution $f_{H_{j}}(\bm{p})$ is
{\setlength\arraycolsep{0pt}
\begin{eqnarray}
  f_{H_{j}}(\bm{p})= N_{h_1h_2h_3} 
 \int d\bm{x}_1d\bm{x}_2d\bm{x}_3 d\bm{p}_1 d\bm{p}_2d\bm{p}_3  f^{(n)}_{h_1h_2h_3}(\bm{x}_1,\bm{x}_2,\bm{x}_3;\bm{p}_1,\bm{p}_2,\bm{p}_3)   
 \mathcal {R}_{H_{j}}(\bm{x}_1,\bm{x}_2,\bm{x}_3;\bm{p}_1,\bm{p}_2,\bm{p}_3,\bm{p}).     \label{eq:fHj3hgeneral1}
\end{eqnarray} }%
$N_{h_1h_2h_3}$ is the number of all possible $h_1h_2h_3$-clusters and it equals to $N_{h_1}N_{h_2}N_{h_3},~N_{h_1}(N_{h_1}-1)N_{h_3}$ for $h_1 \neq h_2 \neq h_3$, $h_1 = h_2 \neq h_3$, respectively.
$f^{(n)}_{h_1h_2h_3}(\bm{x}_1,\bm{x}_2,\bm{x}_3;\bm{p}_1,\bm{p}_2,\bm{p}_3)$ is the normalized three-hadron joint coordinate-momentum distribution,
and $\mathcal {R}_{H_{j}}(\bm{x}_1,\bm{x}_2,\bm{x}_3;\bm{p}_1,\bm{p}_2,\bm{p}_3,\bm{p})$ is the kernel function.

We rewrite the kernel function as
{\setlength\arraycolsep{0pt}
\begin{eqnarray}
  \mathcal {R}_{H_{j}}(\bm{x}_1,\bm{x}_2,\bm{x}_3;\bm{p}_1,\bm{p}_2,\bm{p}_3,\bm{p}) = g_{H_{j}}
	\mathcal {R}_{H_{j}}^{(x,p)}(\bm{x}_1,\bm{x}_2,\bm{x}_3;\bm{p}_1,\bm{p}_2,\bm{p}_3)    \delta(\displaystyle{\sum^3_{i=1}} \bm{p}_i-\bm{p}).   \label{eq:RHj3h}  
\end{eqnarray} }%
The spin degeneracy factor $g_{H_{j}} = (2J_{H_{j}}+1) /[\prod \limits_{i=1}^3(2J_{h_i}+1)]$.
$\mathcal {R}_{H_{j}}^{(x,p)}(\bm{x}_1,\bm{x}_2,\bm{x}_3;\bm{p}_1,\bm{p}_2,\bm{p}_3)$ solving from the Wigner transformation~\cite{Chen:2003ava,Ko:2012lhi,Zhu:2015voa} is
{\setlength\arraycolsep{0pt}
\begin{eqnarray}
  \mathcal {R}^{(x,p)}_{H_{j}}(\bm{x}_1,\bm{x}_2,\bm{x}_3;\bm{p}_1,\bm{p}_2,\bm{p}_3) = 8^2 e^{-\frac{(\bm{x}'_1-\bm{x}'_2)^2}{2\sigma_1^2}} 
e^{-\frac{2(\frac{m_1\bm{x}'_1}{m_1+m_2}+\frac{m_2\bm{x}'_2}{m_1+m_2}-\bm{x}'_3)^2}{3\sigma_2^2}}  e^{-\frac{2\sigma_1^2(m_2\bm{p}'_{1}-m_1\bm{p}'_{2})^2}{(m_1+m_2)^2}} 
	e^{-\frac{3\sigma_2^2[m_3\bm{p}'_{1}+m_3\bm{p}'_{2}-(m_1+m_2)\bm{p}'_{3}]^2} {2(m_1+m_2+m_3)^2}}.~~~~~~      \label{eq:RHj3hxp} 
\end{eqnarray} }%
The superscript `$'$' denotes the hadron coordinate or momentum in the rest frame of the $h_1h_2h_3$-cluster.
The width parameter $\sigma_1=\sqrt{\frac{m_3(m_1+m_2)(m_1+m_2+m_3)} {m_1m_2(m_1+m_2)+m_2m_3(m_2+m_3)+m_3m_1(m_3+m_1)}} R_{H_{j}}$,
and $\sigma_2=\sqrt{\frac{4m_1m_2(m_1+m_2+m_3)^2} {3(m_1+m_2)[m_1m_2(m_1+m_2)+m_2m_3(m_2+m_3)+m_3m_1(m_3+m_1)]}} R_{H_{j}}$,
where $R_{H_{j}}$ is the root-mean-square radius of the $H_{j}$.

Substituting Eqs.~(\ref{eq:RHj3h}) and (\ref{eq:RHj3hxp}) into Eq.~(\ref{eq:fHj3hgeneral1}), we have
{\setlength\arraycolsep{0.2pt}
\begin{eqnarray}
 f_{H_{j}}(\bm{p})=&&8^2N_{h_1h_2h_3} g_{H_{j}} \int d\bm{x}_1d\bm{x}_2d\bm{x}_3 d\bm{p}_1 d\bm{p}_2d\bm{p}_3  e^{-\frac{(\bm{x}'_1-\bm{x}'_2)^2}{2\sigma_1^2}} 
  e^{-\frac{2(\frac{m_1\bm{x}'_1}{m_1+m_2}+\frac{m_2\bm{x}'_2}{m_1+m_2}-\bm{x}'_3)^2}{3\sigma_2^2}}   f^{(n)}_{h_1h_2h_3}(\bm{x}_1,\bm{x}_2,\bm{x}_3;\bm{p}_1,\bm{p}_2,\bm{p}_3)     \nonumber  \\
&& \times  e^{-\frac{2\sigma_1^2(m_2\bm{p}'_{1}-m_1\bm{p}'_{2})^2}{(m_1+m_2)^2}}
 e^{-\frac{3\sigma_2^2[m_3\bm{p}'_{1}+m_3\bm{p}'_{2}-(m_1+m_2)\bm{p}'_{3}]^2} {2(m_1+m_2+m_3)^2}}   \delta(\displaystyle{\sum^3_{i=1}} \bm{p}_i-\bm{p}).      \label{eq:fHj3h}  
\end{eqnarray} }%

Approximating the Gaussian form of the momentum-dependent kernel function to be $\delta$ function form and integrating $\bm{p}_1$, $\bm{p}_2$ and $\bm{p}_3$ from Eq.~(\ref{eq:fHj3h}), we can obtain
{\setlength\arraycolsep{0.2pt}
\begin{eqnarray}
f_{H_{j}}(\bm{p})=&& 8^2 N_{h_1h_2h_3} g_{H_{j}} \int d\bm{x}_1d\bm{x}_2d\bm{x}_3d\bm{p}_1d\bm{p}_2d\bm{p}_3   f^{(n)}_{h_1h_2h_3}(\bm{x}_1,\bm{x}_2,\bm{x}_3;\bm{p}_1,\bm{p}_2,\bm{p}_3) 
   e^{-\frac{(\bm{x}'_1-\bm{x}'_2)^2}{2\sigma_1^2}} e^{-\frac{2(\frac{m_1\bm{x}'_1}{m_1+m_2}+\frac{m_2\bm{x}'_2}{m_1+m_2}-\bm{x}'_3)^2}{3\sigma_2^2}}  \nonumber   \\
		&&\times   (\frac{\sqrt{\pi}}{\sqrt{2} \sigma_1})^3  (1+\frac{m_1}{m_2})^3
		\delta(\bm{p}'_{1}-\frac{m_1}{m_2} \bm{p}'_{2})   (\frac{\sqrt{2\pi} }{\sqrt{3} \sigma_2})^3 (1+\frac{m_1}{m_3}+\frac{m_2}{m_3})^3
		\delta(\bm{p}'_{1}+\bm{p}'_{2}-\frac{m_1+m_2}{m_3} \bm{p}'_{3}) \delta(\displaystyle{\sum^3_{i=1}} \bm{p}_i-\bm{p})          \nonumber   \\	
= && 8^2 N_{h_1h_2h_3} g_{H_{j}} \int d\bm{x}_1d\bm{x}_2d\bm{x}_3d\bm{p}_1d\bm{p}_2d\bm{p}_3  f^{(n)}_{h_1h_2h_3}(\bm{x}_1,\bm{x}_2,\bm{x}_3;\bm{p}_1,\bm{p}_2,\bm{p}_3)
    e^{-\frac{(\bm{x}'_1-\bm{x}'_2)^2}{2\sigma_1^2}}e^{-\frac{2(\frac{m_1\bm{x}'_1}{m_1+m_2}+\frac{m_2\bm{x}'_2}{m_1+m_2}-\bm{x}'_3)^2}{3\sigma_2^2}}  \nonumber   \\
		&& \times 	(\frac{\sqrt{\pi}}{\sqrt{2} \sigma_1})^3  (1+\frac{m_1}{m_2})^3
    	\gamma	\delta(\bm{p}_{1}-\frac{m_1}{m_2} \bm{p}_{2})   (\frac{\sqrt{2\pi}}{\sqrt{3} \sigma_2})^3 (1+\frac{m_1}{m_3}+\frac{m_2}{m_3})^3
		\gamma \delta(\bm{p}_{1}+\bm{p}_{2}-\frac{m_1+m_2}{m_3} \bm{p}_{3})  \delta(\displaystyle{\sum^3_{i=1}} \bm{p}_i-\bm{p})          \nonumber   \\             
=&& 8^2 N_{h_1h_2h_3} g_{H_{j}} \gamma{^2}  (\frac{\pi}{\sqrt{3}\sigma_1\sigma_2})^3 
  \int d\bm{x}_1d\bm{x}_2d\bm{x}_3  f^{(n)}_{h_1h_2h_3}(\bm{x}_1,\bm{x}_2,\bm{x}_3;\frac{m_1 \bm{p}}{m_1+m_2+m_3} , \frac{m_2 \bm{p}}{m_1+m_2+m_3},\frac{m_3 \bm{p}}{m_1+m_2+m_3}) \nonumber   \\
  && \times e^{-\frac{(\bm{x}'_1-\bm{x}'_2)^2}{2\sigma_1^2}} e^{-\frac{2(\frac{m_1\bm{x}'_1}{m_1+m_2}+\frac{m_2\bm{x}'_2}{m_1+m_2}-\bm{x}'_3)^2}{3\sigma_2^2}}. \label{eq:fH3-p}   
\end{eqnarray} }%
		
Changing coordinate variables in Eq.~(\ref{eq:fH3-p}) to be $\bm{Y}= (m_1\bm{x}_1+m_2\bm{x}_2+m_3\bm{x}_3)/(m_1+m_2+m_3)$, $\bm{r}_1= (\bm{x}_1-\bm{x}_2)/\sqrt{2}$ and 
$\bm{r}_2=\sqrt{\frac{2}{3}} (\frac{m_1\bm{x}_1}{m_1+m_2}+\frac{m_2\bm{x}_2}{m_1+m_2}-\bm{x}_3)$ as in Refs.~\cite{Chen:2003ava,Ko:2012lhi,Zhu:2015voa}, we have  
{\setlength\arraycolsep{0pt}
\begin{eqnarray}
f_{H_j}(\bm{p})&=& 8^2 N_{h_1h_2h_3} g_{H_{j}} \gamma{^2}  (\frac{\pi}{\sqrt{3}\sigma_1\sigma_2})^3  \nonumber  \\
 && \times
  \int 3^{3/2}d\bm{Y}d\bm{r_1}d\bm{r_2}  f^{(n)}_{h_1h_2h_3}(\bm{Y},\bm{r}_1,\bm{r}_2;\frac{m_1 \bm{p}}{m_1+m_2+m_3}, \frac{m_2 \bm{p}}{m_1+m_2+m_3}, \frac{m_3 \bm{p}}{m_1+m_2+m_3}) e^{-\frac{\bm{r}_1'^2}{\sigma_1^2}} e^{-\frac{\bm{r}_2'^2}{\sigma_2^2}}.     \label{eq:fH3-Xr}  
\end{eqnarray} }%
We also assume the center-of-mass coordinate in joint distribution is factorized, i.e.,
{\setlength\arraycolsep{0pt}
\begin{eqnarray}
&& 3^{3/2} f^{(n)}_{h_1h_2h_3}(\bm{Y},\bm{r}_1,\bm{r}_2;\frac{m_1  \bm{p}}{m_1+m_2+m_3}, \frac{m_2  \bm{p}}{m_1+m_2+m_3},\frac{m_3 \bm{p}}{m_1+m_2+m_3} ) \nonumber   \\
 = && f^{(n)}_{h_1h_2h_3}(\bm{Y}) f^{(n)}_{h_1h_2h_3}(\bm{r}_1,\bm{r}_2;\frac{m_1  \bm{p}}{m_1+m_2+m_3},\frac{m_2 \bm{p}}{m_1+m_2+m_3}, \frac{m_3  \bm{p}}{m_1+m_2+m_3}).  \label{eq:fh1h2h3-Xr}  
\end{eqnarray} }%
Substituting Eq.~(\ref{eq:fh1h2h3-Xr}) into Eq.~(\ref{eq:fH3-Xr}), we have
{\setlength\arraycolsep{0pt}
\begin{eqnarray}
 f_{H_{j}}(\bm{p})=&& 8^2 N_{h_1h_2h_3} g_{H_{j}} \gamma{^2}  (\frac{\pi}{\sqrt{3}\sigma_1\sigma_2})^3 
  \int d\bm{r_1}d\bm{r_2}  f^{(n)}_{h_1h_2h_3}(\bm{r}_1,\bm{r}_2;\frac{m_1  \bm{p}}{m_1+m_2+m_3}, \frac{m_2 \bm{p}}{m_1+m_2+m_3},  \frac{m_3  \bm{p}}{m_1+m_2+m_3})  \nonumber \\
&& \times   e^{-\frac{\bm{r}_1'^2}{\sigma_1^2}} e^{-\frac{\bm{r}_2'^2}{\sigma_2^2}}.   \label{eq:fH3-r} 
\end{eqnarray} }%
		
Adopting Gaussian form for the relative coordinate distribution~\cite{Mrowczynski:2016xqm,Wang:2020zaw,Kisiel:2014upa,ALICE:2015tra}, we have
{\setlength\arraycolsep{0pt}
\begin{eqnarray}
&&~~~ f^{(n)}_{h_1h_2h_3}(\bm{r}_1,\bm{r}_2;\frac{m_1  \bm{p}}{m_1+m_2+m_3}, \frac{m_2 \bm{p}}{m_1+m_2+m_3},\frac{m_3 \bm{p}}{m_1+m_2+m_3}) \nonumber   \\ 
&&=  \frac{1}{[\pi C_1 R_f^2(\bm{p})]^{3/2}} e^{-\frac{\bm{r}_1^2}{C_1 R_f^2(\bm{p})}} \frac{1}{[\pi C_2 R_f^2(\bm{p})]^{3/2}}  
  e^{-\frac{\bm{r}_2^2}{C_2 R_f^2(\bm{p})}} f^{(n)}_{h_1h_2h_3}(\frac{m_1  \bm{p}}{m_1+m_2+m_3}, \frac{m_2 \bm{p}}{m_1+m_2+m_3},\frac{m_3 \bm{p}}{m_1+m_2+m_3}).  \label{eq:fh1h2h3-rp}    
\end{eqnarray} }%
Comparing relations of $\bm{r}_1$, $\bm{r}_2$ with $\bm{x}_1$, $\bm{x}_2$, $\bm{x}_3$ to that of $\bm{r}$ with $\bm{x}_1$, $\bm{x}_2$ in Sec.~\ref{2hcoHj}, 
we see that $C_1$ is equal to $C_0$ and $C_2$ is $4C_0/3$~\cite{Mrowczynski:2016xqm,Wang:2020zaw,Kisiel:2014upa,ALICE:2015tra}. 
Substituting Eq.~(\ref{eq:fh1h2h3-rp}) into Eq.~(\ref{eq:fH3-r}) and considering the coordinate Lorentz transformation, we integrate from the relative coordinate variables and obtain 
{\setlength\arraycolsep{0pt}
\begin{eqnarray}
f_{H_{j}}(\bm{p}) &=& \frac{8^2 \pi^3 g_{H_{j}} \gamma^2}{3^{3/2} \left[C_1 R_f^2(\bm{p})+\sigma_1^2\right] \sqrt{C_1 [R_f(\bm{p})/\gamma]^2+\sigma_1^2}
\left[C_2 R_f^2(\bm{p})+\sigma_2^2\right] \sqrt{C_2 [R_f(\bm{p})/\gamma]^2+\sigma_2^2}}    \nonumber   \\
		&& \times f_{h_1h_2h_3}(\frac{m_1  \bm{p}}{m_1+m_2+m_3},  \frac{m_2 \bm{p}}{m_1+m_2+m_3},\frac{m_3 \bm{p}}{m_1+m_2+m_3}) .   \label{eq:fHj3h-}
\end{eqnarray} }%
Ignoring correlations between $h_1$, $h_2$ and $h_3$, we have the three-dimensional momentum distribution of $H_j$ as
{\setlength\arraycolsep{0pt}
\begin{eqnarray}
f_{H_{j}}(\bm{p}) &=& \frac{8^2 \pi^3 g_{H_{j}} \gamma^2}{3^{3/2}\left[C_1 R_f^2(\bm{p})+\sigma_1^2\right] \sqrt{C_1 [R_f(\bm{p})/\gamma]^2+\sigma_1^2}
\left[C_2 R_f^2(\bm{p})+\sigma_2^2\right] \sqrt{C_2 [R_f(\bm{p})/\gamma]^2+\sigma_2^2}}     \nonumber   \\
&& \times    f_{h_1}(\frac{m_1\bm{p}}{m_1+m_2+m_3})f_{h_2}(\frac{m_2\bm{p}}{m_1+m_2+m_3})  f_{h_3}(\frac{m_3\bm{p}}{m_1+m_2+m_3}) .     \label{eq:fHj3h-approx}
\end{eqnarray} }%
		
Finally, we have the Lorentz invariant momentum distribution as
{\setlength\arraycolsep{0.2pt}
\begin{eqnarray}
f_{H_j}^{(inv)}(p_{T},y) &=& \frac{8^2 \pi^3 g_{H_{j}} }{3^{3/2}\left[C_1 R_f^2(p_T,y)+\sigma_1^2\right] \sqrt{C_1 [R_f(p_T,y)/\gamma]^2+\sigma_1^2}
\left[C_2 R_f^2(p_T,y)+\sigma_2^2\right] \sqrt{C_2 [R_f(p_T,y)/\gamma]^2+\sigma_2^2}}     \nonumber   \\
&&  \times   \frac{m_{H_j}}{m_1m_2m_3}  f_{h_1}^{(inv)}(\frac{m_1p_{T}}{m_1+m_2+m_3},y)  f_{h_2}^{(inv)}(\frac{m_2p_{T}}{m_1+m_2+m_3},y) f_{h_3}^{(inv)}(\frac{m_3p_{T}}{m_1+m_2+m_3},y).    \label{eq:pt-Hj3h}
\end{eqnarray} }%
\end{widetext}

As a short summary of this section, we want to state that Eqs.~(\ref{eq:pt-Hj2h}) and (\ref{eq:pt-Hj3h}) give: 
(i) the relationships of light (hyper-)nuclei with primordial baryons in momentum space in the laboratory frame,
(ii) effects of different factors on light (hyper-)nucleus production such as the whole hadronic system scale and the size of the formed nuclei.
They can be directly used to calculate the rapidity and $p_T$ distributions of light (hyper-)nuclei as long as the primordial momentum distributions of nucleons and hyperons are given.
What's more, they can be conveniently used to investigate the production correlations of different species of light (hyper-)nuclei.
Their applications at midrapidity (i.e., $y=0$) in heavy-ion collisions at the RHIC will be shown in the following sections.

\section{Transverse momentum spectra of light (hyper-)nuclei} \label{pTresults}

In this section, we use the coalescence model to study the $p_T$ spectra of light and hyper-nuclei in central Au-Au collisions at the RHIC energies.
We first introduce the $p_T$ spectra of nucleons and hyperons obtained from the QCM~\cite{Feng:2022dln}. 
We then investigate the $p_T$ spectra of $d$, $t$, $^3$He, $^3_\Lambda$H, $H(p\Omega^-)$, $H(n\Omega^-)$ and $H(pn\Omega^-)$.
We finally present the average transverse momenta $\langle p_T \rangle$ and the yield rapidity densities $dN/dy$ of different light (hyper-)nuclei.

\subsection{$p_T$ spectra of nucleons and hyperons}  \label{np-pTdistribution}

\begin{figure}[htbp]
\centering
 \includegraphics[width=0.95\linewidth]{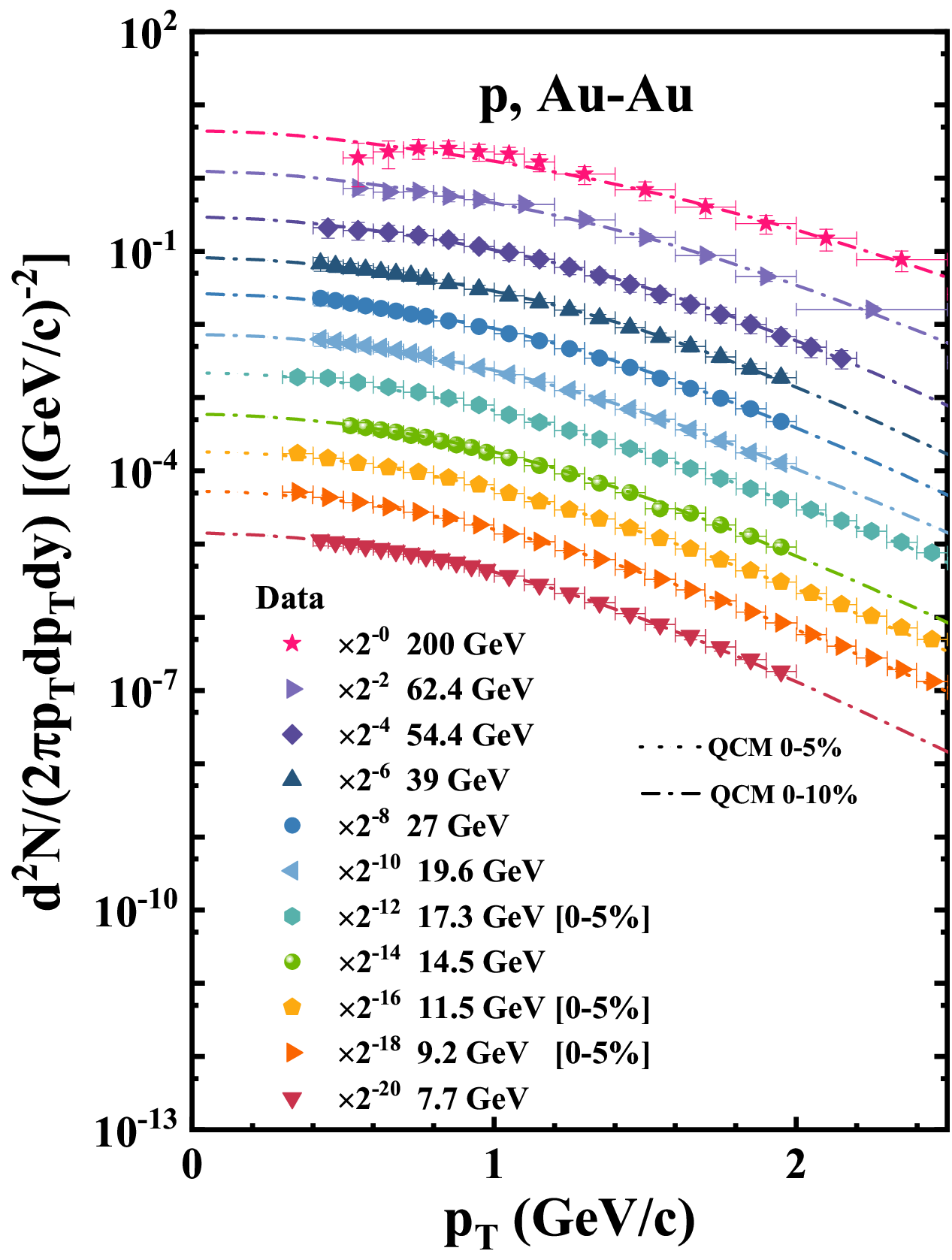}\\
 \caption{Invariant $p_T$ spectra of primordial protons in Au-Au collisions in $0-5$\% centrality at $\sqrt{s_{NN}}=9.2,~11.5,~17.3$ GeV and $0-10$\% centrality at $\sqrt{s_{NN}}=7.7,~14.5,~19.6,~27,~39,~54.4,~62.4,~200$ GeV. Filled symbols are experimental data~\cite{STAR:2022hbp,liu2025measurement}. 
Different lines are results of the QCM~\cite{Feng:2022dln}.}
 \label{fig:ppT-BES}
\end{figure}

\begin{figure}[htbp]
\centering
 \includegraphics[width=0.95\linewidth]{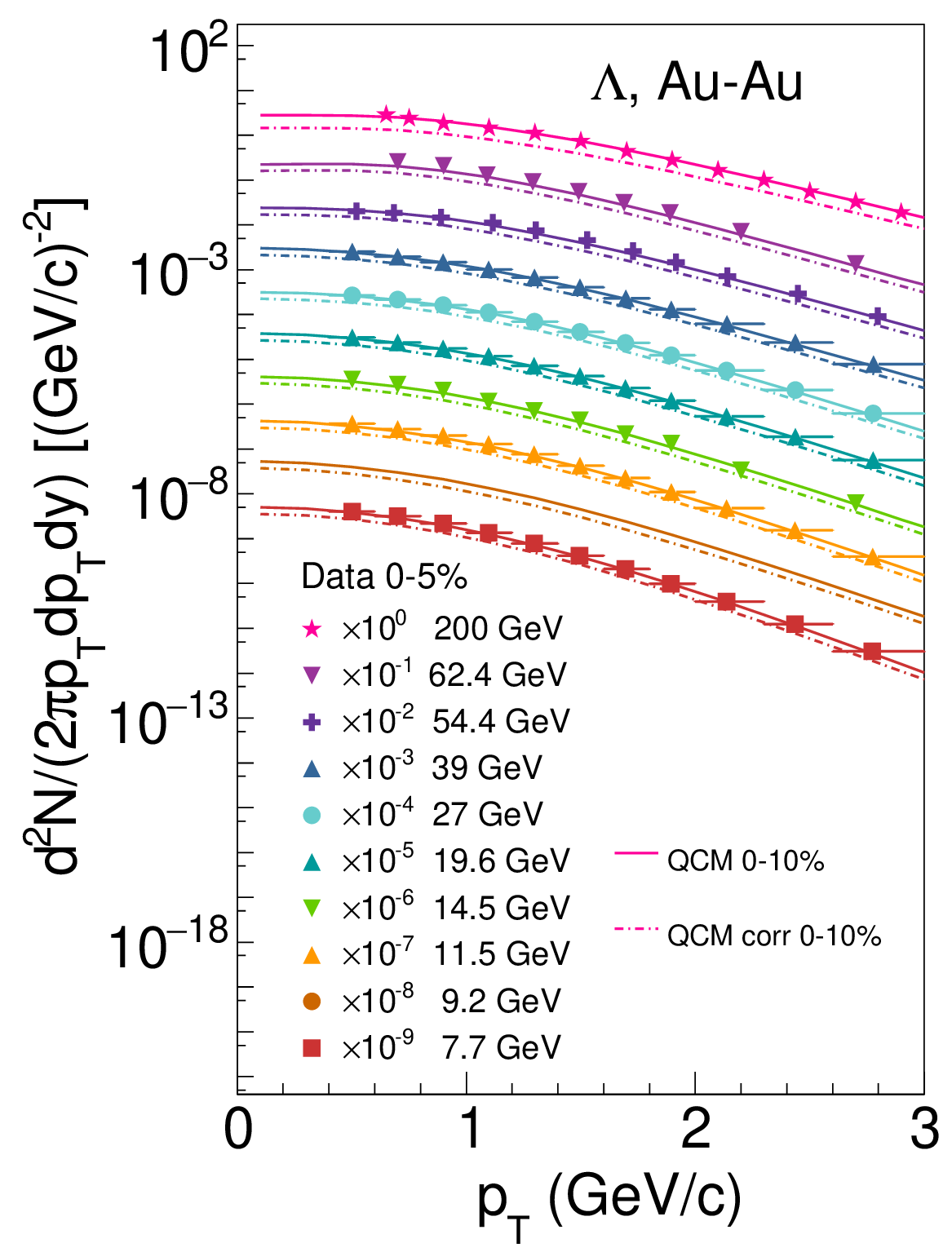}\\
 \caption{Invariant $p_T$ spectra of $\Lambda$ in central Au-Au collisions at $\sqrt{s_{NN}}=7.7,~9.2,~11.5,~14.5,~19.6,~27,~39,~54.4,~62.4$, and 200 GeV. 
Filled symbols are experimental data~\cite{Bairathi:2015kym,STAR:2019bjj,STAR:2010yyv,STAR:2006egk}.
Solid lines are the results of the QCM with $\Sigma^0$ decay, and dashed-dotted lines are those of the QCM corrected contamination from $\Sigma^0$ decay~\cite{Feng:2022dln}.}
 \label{fig:LampT-BES}
\end{figure}

\begin{figure}[htbp]
\centering
 \includegraphics[width=1.\linewidth]{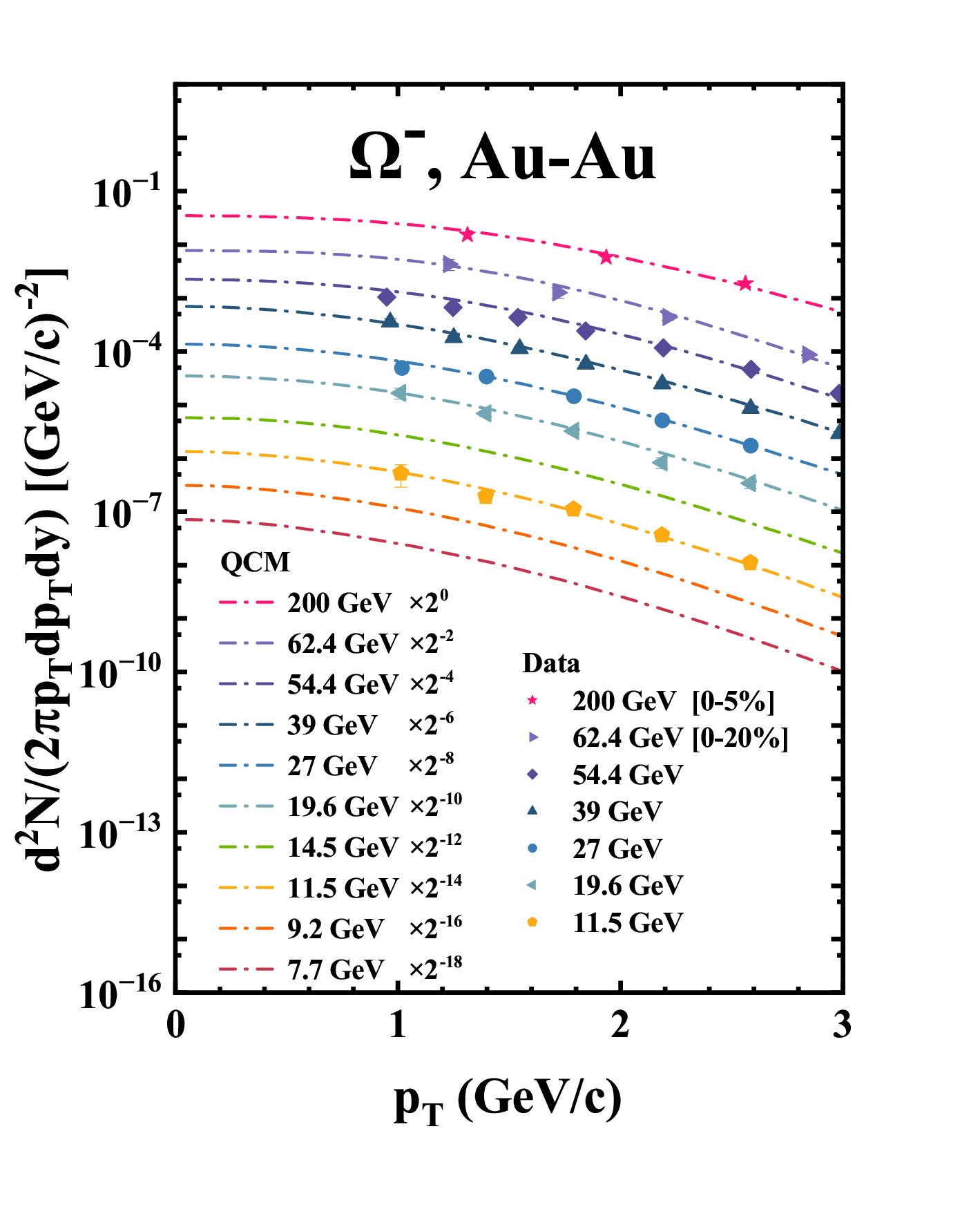}\\
 \caption{Invariant $p_T$ spectra of $\Omega^-$ in central Au-Au collisions at $\sqrt{s_{NN}}=7.7,~9.2,~11.5,~14.5,~19.6,~27,~39,~54.4,~62.4$, and 200 GeV. 
Filled symbols are experimental data~\cite{STAR:2019bjj,STAR:2010yyv,STAR:2006egk}, and different lines are the results of the QCM~\cite{Feng:2022dln}.}
 \label{fig:OmegapT-BES}
\end{figure}

The $p_T$ spectra of primordial nucleons and hyperons are necessary inputs for computing $p_T$ distributions of light nuclei and hyper-nuclei in the coalescence model.
We here use the QCM developed by the Shandong group~\cite{Liang:1991ya,Shao:2004cn,Chang:2023zbe} to get $p_T$ distributions of $p$, $n$, $\Lambda$ and $\Omega^-$.
QCM is developed for describing hadron production exclusively, and so can provide a convenient way to explore production correlations of different kinds of hadrons~\cite{Wang:2012cw,Wang:2013duu,Feng:2025wde}.

Figure \ref{fig:ppT-BES} shows the invariant $p_T$ spectra of primordial protons at midrapidity in central Au-Au collisions at $\sqrt{s_{NN}}=7.7,~9.2,~11.5,~14.5,~17.3,~19.6,~27,~39,~54.4,~62.4$, and 200 GeV.
The spectra in different energies are scaled by different factors for clarity as shown in the figure.
Filled symbols are experimental data in $0-5$\% centrality at $\sqrt{s_{NN}}=9.2,~11.5,~17.3$ GeV from the RHIC BES II of the STAR collaboration and $0-10$\% centrality at $\sqrt{s_{NN}}=7.7,~14.5,~19.6,~27,~39,~54.4,~62.4,~200$ GeV from BES I~\cite{liu2025measurement,STAR:2022hbp}.  
Dotted lines are results of the QCM for the $0-5$\% centrality at $\sqrt{s_{NN}}=9.2,~11.5,~17.3$ GeV and dashed-dotted lines for the $0-10$\% centrality at other collision energies~\cite{Feng:2022dln}.

For the primordial neutrons, they have the same $p_T$ distribution shape with those of the protons at the same collision condition.
But the absolute yield is slightly surplus due to the influence of the net nucleons from the colliding Au nuclei.
$Z_{np}$ is employed to denote the neutron-to-proton yield ratio, and its values are obtained via the QCM to be 1.12, 1.11, 1.10, 1.08, 1.07, 1.07, 1.05, 1.04, 1.03, 1.02, 1.01 in central Au-Au collisions at $\sqrt{s_{NN}}=7.7,~9.2,~11.5,~14.5,~17.3,~19.6,~27,~39,~54.4,~62.4,~200$ GeV, respectively~\cite{Feng:2022dln}.
With the increasing collision energy, the Au stopping power becomes weak and the influence from the net nucleons stopped at midrapidity becomes faint.
This makes $Z_{np}$ decrease with the increasing collision energy.
Actually, $Z_{np}$ decreases to unity up to LHC energies~\cite{Wang:2013pnn,Chang:2023zbe}, which means that $p$ and $n$ are equally created at the midrapidity area.

Figure~\ref{fig:LampT-BES} shows the invariant $p_T$ spectra of $\Lambda$ hyperons at midrapidity in central Au-Au collisions at $\sqrt{s_{NN}}=$ 7.7, 9.2, 11.5, 14.5, 19.6, 27, 39, 54.4, 62.4, and 200 GeV. 
Solid lines are the $0-10$\% centrality results of the QCM excluding contaminations from hyperon weak decays but including those from $\Sigma^0$ electromagnetic decays~\cite{Feng:2022dln}.
Such decay contamination deduction has been adopted by experiments.
Dashed-dotted lines are the $0-10$\% centrality results of the QCM with further $\Sigma^0$ decay corrections, which are just the input primordial $\Lambda$'s in the coalescence model to compute hypertriton production.
As mean lifetime of $\Sigma^0$ is much longer than the hadronic system evolution duration, $\Lambda$'s from $\Sigma^0$ decays cannot participate in the coalescence process to form hyper-nuclei.
Experimental data in $0-5$\% centrality are also plotted with filled symbols for comparison~\cite{Bairathi:2015kym,STAR:2019bjj,STAR:2010yyv,STAR:2006egk}.

Figure~\ref{fig:OmegapT-BES} shows the invariant $p_T$ spectra of $\Omega^-$ hyperons at midrapidity in central Au-Au collisions at $\sqrt{s_{NN}}=$ 7.7, 9.2, 11.5, 14.5, 19.6, 27, 39, 54.4, 62.4, and 200 GeV. 
Dashed-dotted lines are the $0-10$\% centrality results of the QCM~\cite{Feng:2022dln}.
Filled symbols are experimental data in $0-10$\% centrality except $0-5$\% centrality at $\sqrt{s_{NN}}=200$ GeV and $0-20$\% centrality at $\sqrt{s_{NN}}=62.4$ GeV~\cite{STAR:2019bjj,STAR:2010yyv,STAR:2006egk}.
The results of the QCM agree with the available data well and provide necessary inputs for the coalescence model to compute the production of $\Omega-$hypernuclei.

\subsection{$p_T$ spectra of light nuclei and hyper-nuclei}

To compute the $p_T$ spectra of light nuclei and hyper-nuclei, the specific form of $R_{f}$ is necessary.
Considering that the number of charged hadrons is a good proxy for the size of the hadronic system, $R_{f}$ should have a linear dependence on the cube root of the charged hadron number.
In relativistic heavy-ion collisions, it has been found that $R_{f}$ adopted as the femtoscopic radius can describe light nucleus production well~\cite{Scheibl:1998tk}.
If this still holds for hyper-nuclei, $R_{f}$ at different collision energies should generally factorize into a linear dependence on the cube root of the rapidity density of charged hadrons $(dN_{\text{ch}}/dy)^{1/3}$ and a power-law dependence on the transverse mass of the formed light (hyper-)nucleus $H_j$~\cite{ALICE:2015tra}.
In this paper, we use a general form
{\setlength\arraycolsep{0.2pt}
\begin{eqnarray}
R_f=k\times\left(\frac{dN_{\text{ch}}}{dy}\right)^{1/3}\times\left(\sqrt{p_T^2+m^2_{H_j}}\right)^a+b, ~~~   \label{eq:Rf}
\end{eqnarray} }%
where $k$, $a$ and $b$ are free parameters.
In Ref.~\cite{Zhao:2022xkz}, $a$ was set to be 0, giving $R_f$ independent of $p_T$, i.e., the coordinate-momentum factorization.
Here we also take $a=0$.
Values of $k$ and $b$ in central Au-Au collisions at the RHIC are taken to be (0.515, 0.500) for nuclei with atomic mass number $A=2$ and (0.500, 0.300) for nuclei with $A=3$.
This means slightly smaller $R_f$ of the hadronic system at $A=3$ nucleus freeze-out than that at $A=2$ nucleus freeze-out, which further indicates nuclei with $A=3$ freeze-out earlier.
The $^{3}_{\Lambda}$H is an exception, and it will be discussed later. 

\begin{figure}[htbp]
\centering
 \includegraphics[width=0.95\linewidth]{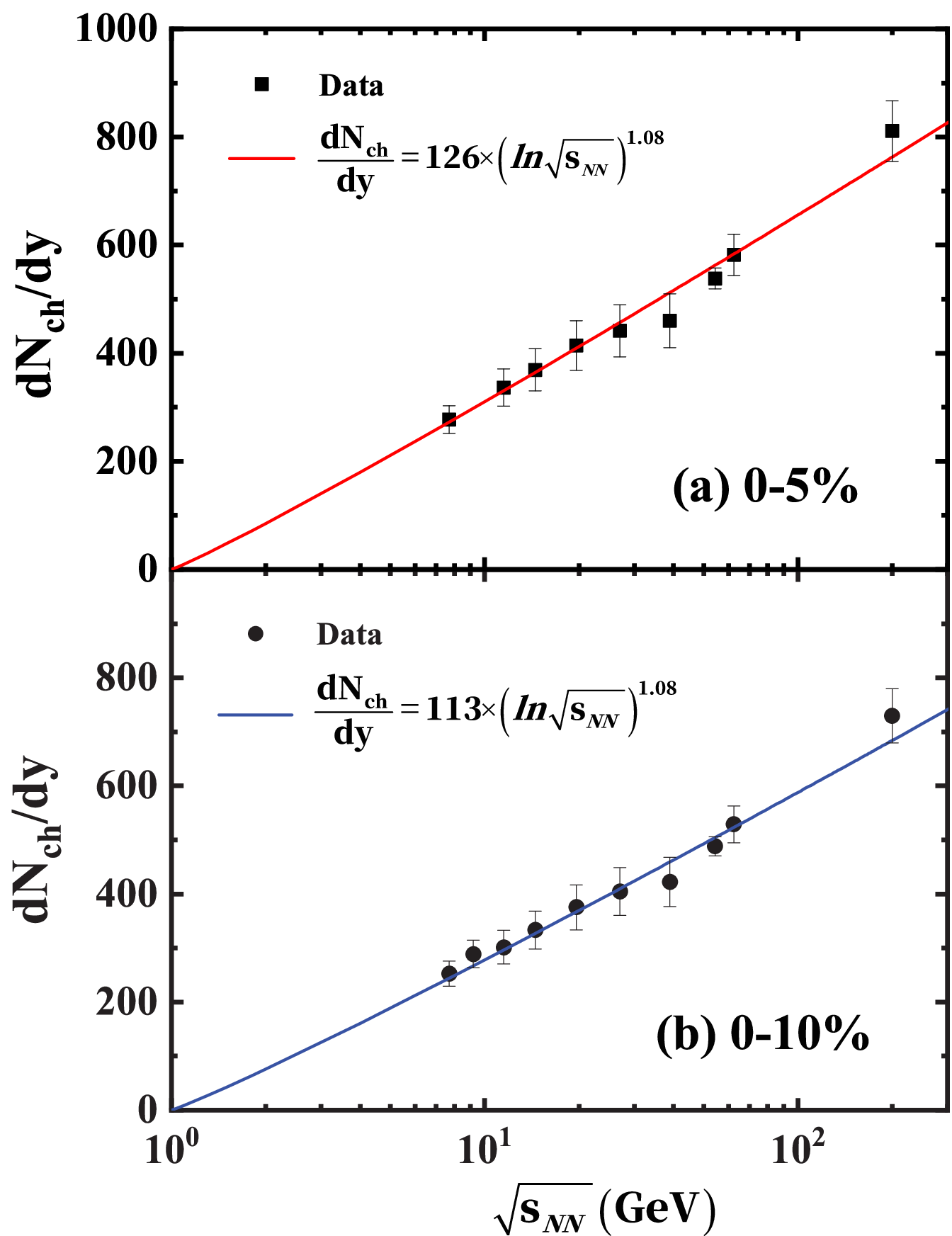}\\
 \caption{Rapidity density of charged hadrons $dN_{\text{ch}}/dy$ as a function of collision energy $\sqrt{s_{NN}}$.
Filled symbols are experimental data~\cite{STAR:2017sal,STAR:2009sxc,STAR:2019vcp,Gopal:2022zgm,STAR:2008med}, and solid lines are parameterized functions.}
 \label{fig:dNchdy-BES}
\end{figure}

We plot the experimental data of $dN_{\text{ch}}/dy$~\cite{STAR:2017sal,STAR:2009sxc,STAR:2019vcp,Gopal:2022zgm,STAR:2008med} in Au-Au collisions in $0-5$\% centrality in Fig.~\ref{fig:dNchdy-BES} (a) and those in $0-10$\% centrality in Fig.~\ref{fig:dNchdy-BES} (b).
Solid lines are the parameterized functions $dN_{\text{ch}}/dy=126\times(\ln\sqrt{s_{NN}})^{1.08}$ and $dN_{\text{ch}}/dy=113\times(\ln\sqrt{s_{NN}})^{1.08}$. 
With these parameterized $dN_{\text{ch}}/dy$ as a function of $\sqrt{s_{NN}}$, we can get $R_f$ at different collision energies, and then compute $p_T$ spectra of different light nuclei and hyper-nuclei.

\begin{figure*}[htbp]
	\centering
	\includegraphics[width=0.9\linewidth]{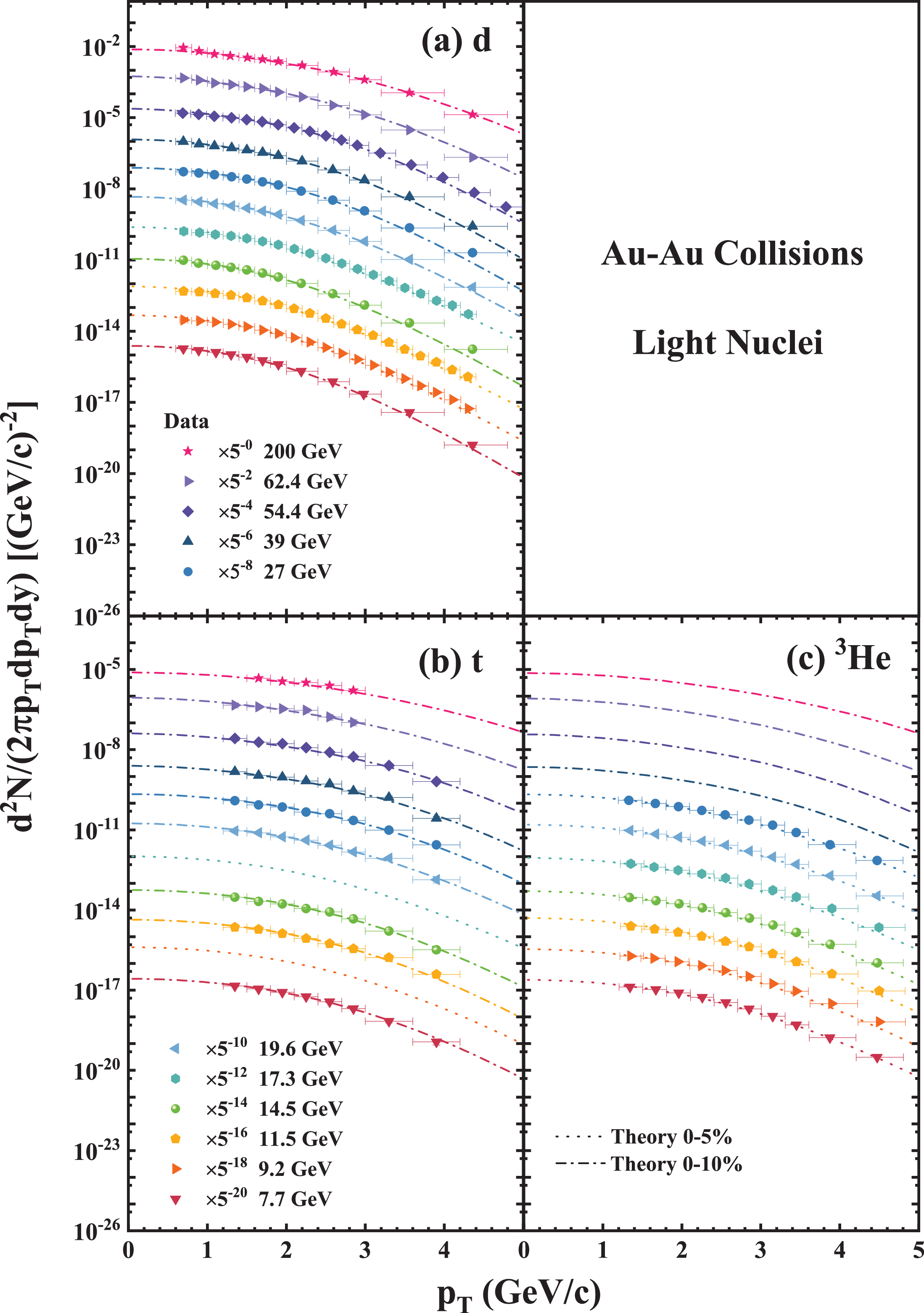}\\
	\caption{Invariant $p_T$ spectra of (a) $d$, (b) $t$, and (c) $^3$He in central Au-Au collisions at $\sqrt{s_{NN}}=7.7$, 9.2, 11.5, 14.5, 19.6, 27, 39, 54.4, 62.4 and 200 GeV. Filled symbols are experimental data~\cite{liu2025measurement,STAR:2022hbp,STAR:2019sjh}. Different lines are theoretical results.}
	\label{fig:dtHe3}
\end{figure*}

With Eqs.~(\ref{eq:pt-Hj2h}) and (\ref{eq:pt-Hj3h}), we first calculate the $p_T$ spectra of $d$, $t$, and $^3$He in central Au-Au collisions at $\sqrt{s_{NN}}=$ 7.7, 9.2, 11.5, 14.5, 17.3, 19.6, 27, 39, 54.4, 62.4 and 200 GeV.
Symbols with error bars in Fig.~\ref{fig:dtHe3} are experimental data from the STAR collaboration~\cite{liu2025measurement,STAR:2022hbp,STAR:2019sjh}. 
Dotted lines are results of the coalescence model for the $0-5$\% centrality and dashed-dotted lines for the $0-10$\% centrality, respectively.
From Fig.~\ref{fig:dtHe3}, one can see the nucleon coalescence can well reproduce the available data of light nuclei at energies from 7.7 to 200 GeV in central Au-Au collisions.

\begin{table*}[htbp]
	\centering
	\caption{Yield rapidity densities $dN/dy$ and average transverse momenta $\langle p_T\rangle$ of $d$, $t$, and $^3\text{He}$ at midrapidity in central Au-Au collisions at $\sqrt{s_{NN}}=7.7$, 9.2, 11.5, 14.5, 17.3, 19.6, 27, 39, 54.4, 62.4 and 200 GeV. The third and fifth columns are experimental data~\cite{liu2025measurement,STAR:2022hbp,STAR:2019sjh}, and the fourth and sixth columns are theoretical results.}    \label{table:dNdypt-dHe3t}
	\resizebox{0.8\textwidth}{!}{
	\begin{tabular}{ccccccc}
		\toprule
		&                          & \multicolumn{2}{c}{$dN/dy$ ($\times10^{-2}$)}                                    &  & $\langle p_T\rangle$ (GeV/c) &         \\ \cline{3-4} \cline{6-7} 
		\multirow{-2}{*}{Light nuclei} & \multirow{-2}{*}{$\sqrt{s_{NN}}$ (GeV)} & Data                    & Theory                             &  & Data                           & Theory          \\ \hline
		& 7.7       &  $140.985 \pm 0.40729 \pm 10.9655 $     &    138.566              & &  $ 1.2746 \pm 0.0006 \pm	0.0336  $         & 1.216     \\
		& 9.2 &$103.5441\pm0.0496\pm5.0495$ &108.523& &$ 1.2953\pm0.0005\pm0.0312$  &1.224   \\
		& 11.5&$69.03 \pm 0.0296 \pm 3.3153$&71.962 & &$ 1.3057\pm0.0007 \pm0.0437$ &1.233   \\
		& 14.5& $43.1475 \pm 0.0941 \pm 3.138 $       & 42.087                &  &  $1.2912 \pm	0.0005 \pm	0.0340  $  & 1.230     \\
		& 17.3&$35.6694\pm0.0192\pm1.6801$  &36.826& &$1.3148\pm	0.0006\pm0.0312$ &1.243   \\
		& 19.6                   & $27.4488 \pm 0.0627 \pm 2.0391 $      & 29.073                &  &  $1.3165 \pm	0.0004 \pm	0.0343  $  & 1.267     \\
		& 27                     & $18.4428 \pm 0.0351 \pm 1.28 $        & 19.826                &  &  $1.3403 \pm	0.0006 \pm	0.0325  $  & 1.287     \\
		& 39                     & $12.7332 \pm 0.0175 \pm 0.952 $       & 12.776                &  &  $1.3762 \pm 0.0019 \pm 0.09 $      & 1.320     \\
		& 54.4                   &$10.256\pm	0.007\pm	0.6$         & 10.274                &  & $---$                                  & 1.351     \\
		& 62.4                   &$ 9.7458 \pm 0.024 \pm 0.64$           &10.029                &  &$ 1.4067 \pm 0.0035 \pm 0.1  $       & 1.405     \\
		\multirow{-13}{*}{${d}$}            
		& 200                    & $7.3238 \pm 0.0098 \pm 0.537 $        & 6.875                &  & $1.5412 \pm 0.0021 \pm 0.1$  & 1.533           \\  
  \hline
		& 7.7        &$ 2.31 \pm 0.05 \pm 0.31   $         & 2.541                   &  &$1.6297\pm0.0343\pm0.0947$                            & 1.553           \\
		& 9.2        & $---$                         & 1.553&  & $---$   	                    	& 1.558    \\
		& 11.5       &$ 0.604 \pm 0.013 \pm 0.06   $       & 0.672                  &  &$1.6794\pm0.035\pm0.0715$                            & 1.565           \\
		& 14.5       & $0.301 \pm 0.007 \pm 0.04   $       & 0.344                 &  &$1.6666\pm0.0355\pm0.0911$                           & 1.568           \\
		& 17.3       & $---$                         & 0.251&  & $---$                         	& 1.572           \\
		& 19.6       & $0.17 \pm 0.004 \pm 0.013  $        & 0.179                 &  &$1.705\pm0.0419\pm0.0556$                          & 1.614           \\
		& 27         &$ 0.0859 \pm 0.0022 \pm 0.007   $    & 0.0915                &  &$1.7457\pm 0.0431\pm0.0627$                           & 1.642           \\
		& 39         &$ 0.043 \pm 0.0009 \pm 0.005   $     & 0.0428                &  &$1.7831\pm 0.0372\pm 0.0436$                           & 1.681          \\
		& 54.4       &$ 0.02941 \pm 0.00032 \pm 0.0019 $   & 0.0284                &  &$1.8096\pm 0.0251 \pm0.1004$                            & 1.716           \\
		& 62.4       &$ 0.024 \pm 0.001 \pm 0.0027 $       & 0.0259                &  &$1.8002\pm 0.0743\pm 0.0535$                          & 1.770           \\
		\multirow{-13}{*}{${t}$}            
		& 200        & $0.0118 \pm 0.0003 \pm 0.0016 $     & 0.0116                &  &$1.9374\pm 0.0452\pm0.1274$                          & 1.936           \\ 
  \hline
		& 7.7            &$2.5084 \pm	0.0062 \pm	0.2565$       &2.289                  &  & $1.6347 \pm	0.0038 \pm	0.0532  $        & 1.547      \\
		& 9.2            &$1.4133\pm	0.0055\pm0.1056$          &1.293  &  & $1.6580\pm0.0042\pm0.0481$               &1.553 \\
		& 11.5           &$0.7373 \pm	0.0032 \pm	0.0432$       &0.657 &  & $1.6700\pm	0.0047 \pm	0.0492  $        & 1.560     \\
		& 14.5           &$0.3518 \pm	0.0014 \pm  0.0379$       &0.314               &  & $1.6952 \pm	0.0059 \pm	0.0515  $        & 1.563      \\
		& 17.3           &$0.2415\pm0.0013\pm0.0198$             &0.219 &  &$1.7065\pm0.0081\pm0.0469$                &1.567   \\
		& 19.6           &$0.1788 \pm	0.0007 \pm	0.0191$      & 0.162                &  & $1.7189 \pm	0.0059 \pm	0.0555  $        & 1.609      \\
		& 27             &$0.1007 \pm	0.0006 \pm	0.0089 $      &0.0898                  &  & $1.7463 \pm	0.0094 \pm	0.0480  $        & 1.637      \\
		& 39             & $---$                                   & 0.0386 &  & $---$                                       & 1.676      \\
		& 54.4           & $---$                                    & 0.0260 &  & $---$                                      & 1.710      \\
		& 62.4           & $---$                                   & 0.0239  &  & $---$                                       & 1.764      \\
		\multirow{-13}{*}{$^3$He}        
		& 200                      & $---$                                   & 0.0109 &  & $---$                                          & 1.929      \\  
		\hline\hline
	\end{tabular}
}
\end{table*}

Table \ref{table:dNdypt-dHe3t} shows the yield rapidity densities $dN/dy$ and the average transverse momenta $\langle p_T\rangle$ of $d$, $t$, and $^3\mathrm{He}$ at midrapidity in central Au-Au collisions at $\sqrt{s_{NN}}=7.7,~9.2,~11.5,~14.5,~17.3,~19.6,~27,~39,~54.4,~62.4$ and 200 GeV.
Experimental data in the third and fifth columns are from Refs.~\cite{liu2025measurement,STAR:2022hbp,STAR:2019sjh}.
Our theoretical results are put in the fourth and sixth columns.
A clear decreasing trend of $dN/dy$ for light nuclei as a function of collision energy is observed.
This mainly results from the decreasing nucleons with the increasing collision energy.
For $\langle p_T\rangle$, an increasing trend is exhibited as a function of collision energy.
This is because that in higher collision energies more energy is deposited in the midrapidity region and collective evolution exists longer.

We then study the production of $^{3}_{\Lambda}$H, which has a very loosely-bound structure and a relatively large size compared to light nuclei due to its much smaller binding energy.
It would be easily destroyed after its formation from freeze-out nucleons and $\Lambda$'s. As a result, the $^3_\Lambda$H is more likely to be produced later than other composite nuclei. 
Therefore, the previously formed deuterons have chance to capture $\Lambda$ to form $^3_\Lambda$H, i.e., $d+\Lambda \rightarrow ^3_{\Lambda}$H, besides the coalescence process $p+n+\Lambda \rightarrow ^3_{\Lambda}$H.
Here we include these two different coalescence contributions.
Based on the above discussion, the hadronic system $R_f$ at $^3_\Lambda$H freeze-out should approach to that at $d$ freeze-out rather than $t$ freeze-out. 
So we use $(0.515,0.500)$ for $k$ and $b$ in Eq.~(\ref{eq:Rf}) to compute $^3_\Lambda$H production.
The $^{3}_{\Lambda}$H root-mean-square radius $R_{^{3}_{\Lambda}\text{H}}$ is adopted to be 4.9 fm~\cite{Nemura:1999qp}.
As the internal structure of $^{3}_{\Lambda}$H is unclear, we adopt two different cases.
One is a spherical shape like the $t$, and the other is a halo structure with a $d$ core encircled by a $\Lambda$.
In the case of the spherical structure, $\sigma_1=\sqrt{\frac{m_{\Lambda}(m_p+m_n)(m_p+m_n+m_{\Lambda})} {m_pm_n(m_p+m_n)+m_nm_{\Lambda}(m_n+m_{\Lambda})+m_{\Lambda}m_p(m_{\Lambda}+m_p)}} R_{^{3}_{\Lambda}\text{H}}$ and $\sigma_2$ $=\sqrt{\frac{4m_pm_n(m_p+m_n+m_{\Lambda})^2} {3(m_p+m_n)[m_pm_n(m_p+m_n)+m_nm_{\Lambda}(m_n+m_{\Lambda})+m_{\Lambda}m_p(m_{\Lambda}+m_p)]}} R_{^{3}_{\Lambda}\text{H}}$.
In the case of the halo structure, $\sigma_1=\sqrt{\frac{2(m_p+m_n)^2}{3(m_p^2+m_n^2)}} R_d$ and $\sigma_2=\sqrt{\frac{2(m_d+m_{\Lambda})^2}{9(m_d^2+m_{\Lambda}^2)}} r_{\Lambda d}$.
The $\Lambda-d$ distance $r_{\Lambda d}$ is evaluated via $r_{\Lambda d}=\sqrt{\hbar^2/(4\mu B_{\Lambda})}$~\cite{Bertulani:2022vad},
where $\mu$ is the reduced mass.
The binding energy $B_{\Lambda}$ here is adopted to be 102 keV measured by the ALICE~\cite{ALICE:2022sco}, 410 keV measured by the STAR~\cite{STAR:2019wjm}, and the world averaged value 148 keV, respectively, to execute calculations.

\begin{table*}[htbp]
	\centering
	\caption{Yield rapidity densities $dN/dy$ and average transverse momenta $\langle p_T\rangle$ of $^3_{\Lambda}\text{H}$ at midrapidity in central Au-Au collisions at $\sqrt{s_{NN}}=7.7,~9.2,~11.5,~14.5,~19.6,~27,~39,~54.4,~62.4$ and 200 GeV. Experimental data are from Ref.~\cite{Li:2025jaz}.}    
	\label{table:dNdypt-hyt}
	\resizebox{1\textwidth}{!}{
		\begin{tabular}{cccccccccccccccccccccccccccccc}
			\hline\hline
			\multirow{4}{*}{$\sqrt{s_{NN}}$ (GeV)} &  & \multicolumn{13}{c}{dN/dy ($\times 10^{-4}$)} &  & \multicolumn{13}{c}{$\langle p_T \rangle$ (GeV/c)} \\ \cline{3-15} \cline{17-29} 
			&  & \multicolumn{11}{c}{Theory} &  & \multirow{3}{*}{Data} &  & \multicolumn{11}{c}{Theory} &  & \multirow{3}{*}{Data} \\ \cline{3-13} \cline{17-27}
			&  & I &  & \multicolumn{5}{c}{II halo} &  & III &  & \multirow{2}{*}{Total} &  &  &  & I &  & \multicolumn{5}{c}{II} &  & III &  & \multirow{2}{*}{Total} &  &  \\
\cline{5-9}   \cline{19-23}
			&  &spherical &  &102 keV &  &148 keV &  &410 keV &  & d+$\Lambda$ &  &  &  &  &  &spherical &  &102 keV &  &148 keV &  &410 keV &  & d+$\Lambda$ &  &  &  &  \\ \cline{1-15} \cline{17-29} 
			7.7 &  & 7.274 &  & 7.282 &  & 10.071 &  & 18.899 &  & 9.895 &  & 28.794 &  & 37.9$\pm$5.1$\pm$5.8 &  & 1.457 &  & 1.465 &  & 1.472 &  & 1.491 &  & 1.471 &  & 1.484 &  & 1.595$\pm$0.076$\pm$0.160 \\
			9.2 &  & 4.860 &  & 4.839 &  & 6.647 &  & 12.253 &  & 6.534 &  & 18.786 &  &$---$  &  & 1.482 &  & 1.489 &  & 1.496 &  & 1.516 &  & 1.496 &  & 1.509 &  &$---$  \\
			11.5 &  & 2.664 &  & 2.636 &  & 3.592 &  & 6.486 &  & 3.533 &  & 10.018 &  & 12.4$\pm$1.8$\pm$2.1 &  & 1.498 &  & 1.505 &  & 1.512 &  & 1.532 &  & 1.512 &  & 1.525 &  & 1.513$\pm$0.068$\pm$0.103 \\
			14.5 &  & 1.670 &  & 1.643 &  & 2.222 &  & 3.933 &  & 2.186 &  & 6.120 &  & 5.42$\pm$0.73$\pm$0.95 &  & 1.505 &  & 1.512 &  & 1.519 &  & 1.539 &  & 1.519 &  & 1.532 &  & 1.659$\pm$0.073$\pm$0.243 \\
			19.6 &  & 1.039 &  & 1.017 &  & 1.363 &  & 2.362 &  & 1.342 &  & 3.704 &  & 4.31$\pm$0.59$\pm$0.69 &  & 1.554 &  & 1.561 &  & 1.569 &  & 1.590 &  & 1.568 &  & 1.582 &  & 1.612$\pm$0.075$\pm$0.109 \\
			27 &  & 0.602 &  & 0.587 &  & 0.780 &  & 1.324 &  & 0.768 &  & 2.092 &  & 2.26$\pm$0.30$\pm$0.31 &  & 1.591 &  & 1.599 &  & 1.607 &  & 1.630 &  & 1.606 &  & 1.621 &  & 1.739$\pm$0.150$\pm$0.177 \\
			39 &  & 0.340 &  & 0.330 &  & 0.435 &  & 0.722 &  & 0.428 &  & 1.150 &  &$---$  &  & 1.618 &  & 1.625 &  & 1.633 &  & 1.657 &  & 1.633 &  & 1.648 &  &$---$  \\
			54.4 &  & 0.240 &  & 0.232 &  & 0.303 &  & 0.495 &  & 0.299 &  & 0.794 &  &$---$  &  & 1.654 &  & 1.662 &  & 1.671 &  & 1.696 &  & 1.670 &  & 1.686 &  &$---$  \\
			62.4 &  & 0.233 &  & 0.226 &  & 0.294 &  & 0.478 &  & 0.290 &  & 0.768 &  &$---$  &  & 1.688 &  & 1.695 &  & 1.705 &  & 1.732 &  & 1.705 &  & 1.722 &  &$---$  \\
			200 &  & 0.130 &  & 0.125 &  & 0.159 &  & 0.247 &  & 0.157 &  & 0.404 &  &$---$  &  & 1.840 &  & 1.848 &  & 1.860 &  & 1.890 &  & 1.859 &  & 1.878 &  &$---$  \\
			\hline\hline
		\end{tabular}
	}
\end{table*}

Table \ref{table:dNdypt-hyt} shows the yield rapidity densities $dN/dy$ and the average transverse momenta $\langle p_T\rangle$ of $^3_{\Lambda}\text{H}$ at midrapidity in central Au-Au collisions at $\sqrt{s_{NN}}=7.7,~9.2,~11.5,~14.5,~19.6,~27,~39,~54.4,~62.4$ and 200 GeV. Experimental data in the eighth and fifteenth columns are from Ref.~\cite{Li:2025jaz}.
The second and the ninth columns are the theoretical results of three-body coalescence in the case of the spherical internal structure of the $^3_{\Lambda}\text{H}$.
The third, fourth and fifth columns are the theoretical $dN/dy$ of three-body coalescence in the case of the halo structure with $B_{\Lambda}=102,~148,~410$ keV, respectively.
The tenth, eleventh and twelfth columns are the corresponding theoretical $\langle p_T\rangle$ with the halo structure.
The sixth and the thirteenth columns are the theoretical results of $d+\Lambda$ two-body coalescence.
A globle decreasing trend of $dN/dy$ while a clear increasing behavior of $\langle p_T\rangle$ from $\sqrt{s_{NN}}=7.7$ GeV to 200 GeV are observed.
This is the same as that of light nuclei.
This is mainly due to the stronger nuclear transparency and more deposited energy in the midrapidity area as a function of collision energy. 
   
From the results of $\langle p_T\rangle$ in Table \ref{table:dNdypt-hyt}, one can see different internal structures of $^3_{\Lambda}\text{H}$ affect its $\langle p_T\rangle$ very slightly.
Our results for all different structures can reproduce the current data within experimental errors. 
For $dN/dy$, theoretical results are very different with different structures.
Results with a spherical structure are comparable to those with a halo structure at $B_{\Lambda}=102$ keV, but obviously lower than those at $B_{\Lambda}=148$ and 410 keV.
This is because larger $B_{\Lambda}$ means tighter binding and relatively smaller size for $^3_{\Lambda}\text{H}$.
Then the production suppression effect from the $^3_{\Lambda}\text{H}$ size becomes relatively weak. 
This leads to an increase of $dN/dy$ with the increasing $B_{\Lambda}$.
The total results at the seventh and fourteenth columns denote the summation of three-body coalescence with a halo $B_{\Lambda}=410$ keV and two-body coalescence,
which agree with the experimental data of both $dN/dy$ and $\langle p_T\rangle$.
Results for both spherical structure and halo with either $B_{\Lambda}=102$ or 148 keV, plus two-body coalescence underestimate the current measured $dN/dy$.
This indicates that our calculations favor that $^3_{\Lambda}\text{H}$ has a halo structure with $B_{\Lambda}=410$ keV.
Very coincidentally, the yield ratios about $^3_{\Lambda}\text{H}$ recently-measured in isobar collisions also support its halo structure with $B_{\Lambda}=410$ keV~\cite{Li:2025awj}.
Future precise measurement for the productivity of $^3_{\Lambda}\text{H}$ can help to further probe its internal construction.
 
\begin{figure}[htbp]
	\centering
	\includegraphics[width=1\linewidth]{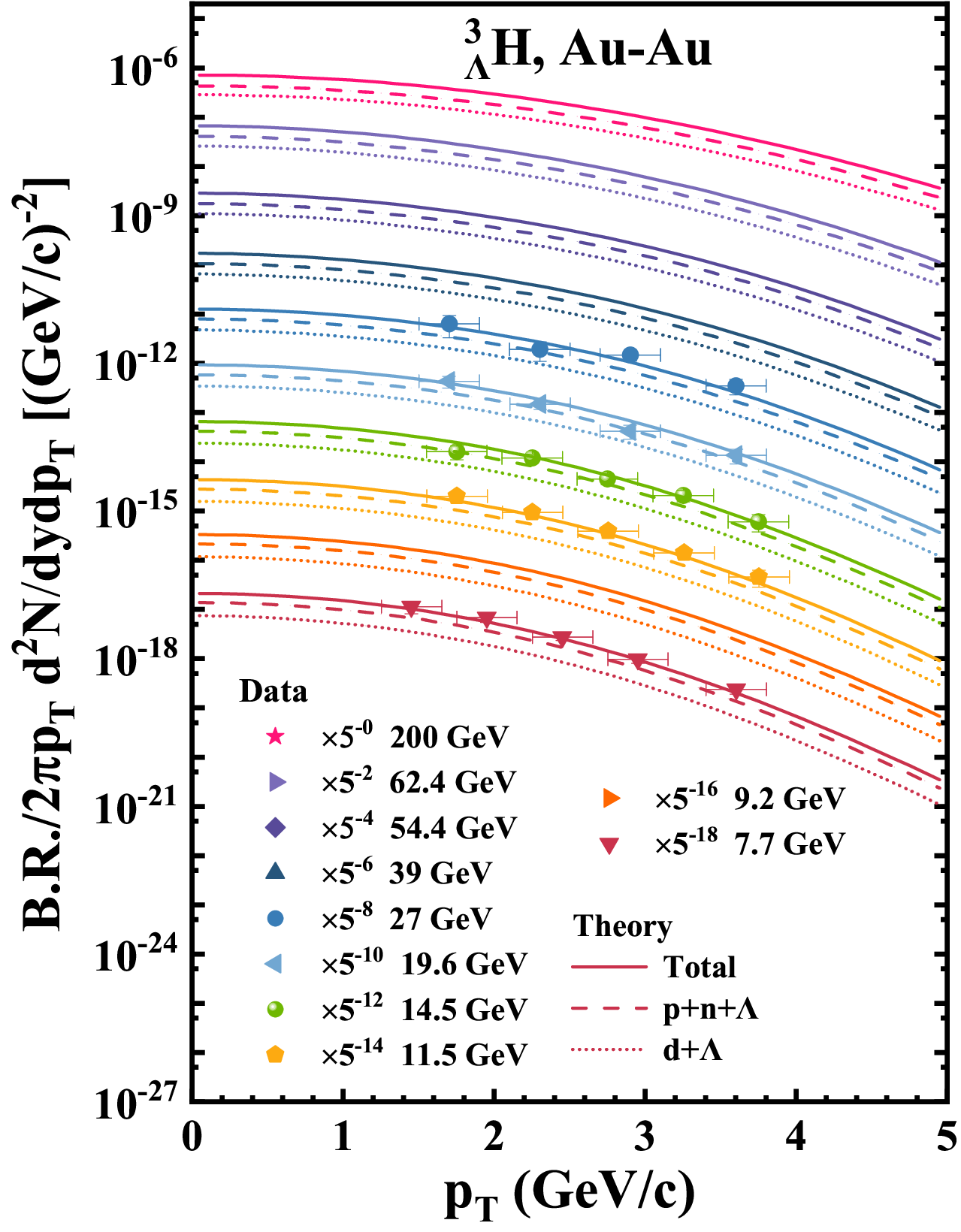}\\
	\caption{Invariant $p_T$ spectra times B.R. of $^3_\Lambda$H in $0-10\%$ centrality of Au-Au collisions at $\sqrt{s_{NN}}=7.7$, 9.2, 11.5, 14.5, 19.6, 27, 39, 54.4, 62.4 and 200 GeV. Filled symbols are experimental data~\cite{Li:2025jaz}. Dashed lines and dotted lines are the theoretical results of $p+n+\Lambda$ coalescence and $d+\Lambda$ coalescence, respectively. Solid lines are the total results of these two coalescence processes.}
	\label{fig:hyt}
\end{figure}

Figure~\ref{fig:hyt} shows the invariant $p_T$ spectra multiplied by the branching ratio (B.R.) of $^{3}_{\Lambda}$H to $^{3}$He in $0-10\%$ centrality of Au-Au collisions at $\sqrt{s_{NN}}=7.7$, 9.2, 11.5, 14.5, 19.6, 27, 39, 54.4, 62.4 and 200 GeV.
Filled symbols with error bars are experimental data~\cite{Li:2025jaz}.
Dashed lines are the theoretical results of the three-body coalescence with a halo structure at $B_{\Lambda}=410$ keV, and dotted lines are those for two-body coalescence.
Solid lines are the total theoretical results of $p+n+\Lambda$ coalescence and $d+\Lambda$ coalescence, which reproduce the available data at $\sqrt{s_{NN}}=7.7$, 11.5, 14.5, 19.6 and 27 GeV well.
Our predictions at other collision energies can provide theoretical references for future experimental measurements.

At the end of this section, we study the production of $\Omega-$hypernuclei $H(p\Omega^-)$, $H(n\Omega^-)$ and $H(pn\Omega^-)$.
The nucleon$-\Omega$ dibaryon states in the S-wave and spin-2 channel have been proposed as interesting candidates for the $d$-like states~\cite{Clement:2016vnl,Pu:2024kfh},
and the $H(pn\Omega^-)$ with maximal spin-$\frac{5}{2}$ is proposed to be a promising partner of the $t$ with multi-strangeness flavor quantum number~\cite{Garcilazo:2019igo}.
Both the STAR and the ALICE have measured the $p-\Omega$ correlation functions, and their results favor the existence of $p\Omega^-$ bound state~\cite{STAR:2018uho,ALICE:2020mfd}.
The HAL QCD collaboration has reported the root-mean-square radius of $H(p\Omega^-)$ is about 3.24 fm and that of $H(n\Omega^-)$ is 3.77 fm~\cite{HALQCD:2018qyu}.
But for $H(pn\Omega^-)$, its root-mean-square radius $R_{H(pn\Omega^-)}$ is undetermined theoretically, and there is no corresponding experimental measurement currently.
We in this work adopt $R_{H(pn\Omega^-)}=2.0$ fm, a typical value for nuclei with several MeV binding energies, to give predictions.

\begin{figure*}[htbp]
	\centering
	\includegraphics[width=0.9\linewidth]{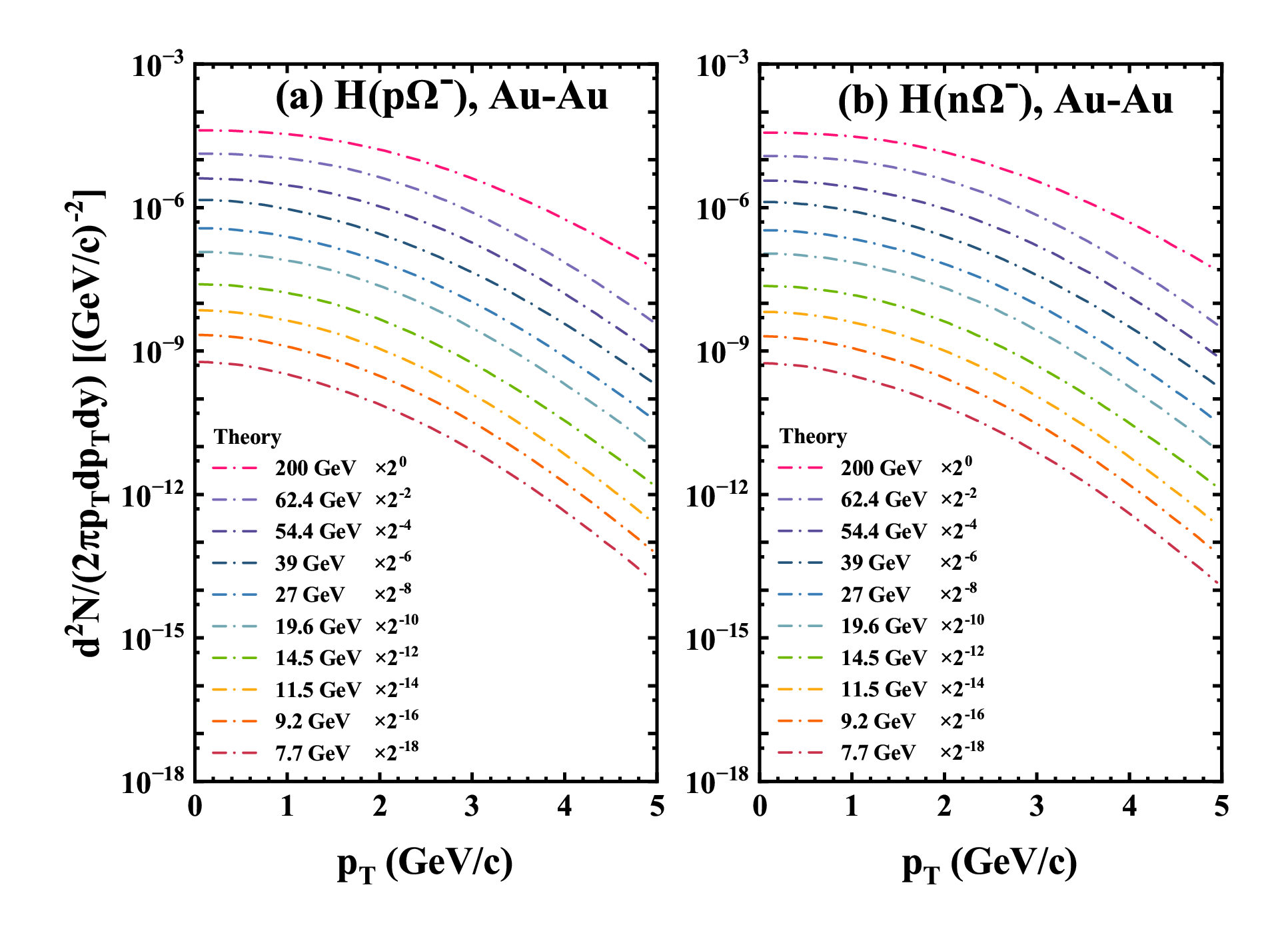}\\
	\caption{Predictions of invariant $p_T$ spectra of (a) $H(p\Omega^-)$ and (b) $H(n\Omega^-)$ in $0-10\%$ centrality of Au-Au collisions at $\sqrt{s_{NN}}$ = 7.7, 9.2, 11.5, 14.5, 19.6, 27, 39, 54.4, 62.4 and 200 GeV. }
	\label{fig:pOnO-pt}
\end{figure*}

\begin{figure}[htbp]
	\centering
	\includegraphics[width=0.99\linewidth]{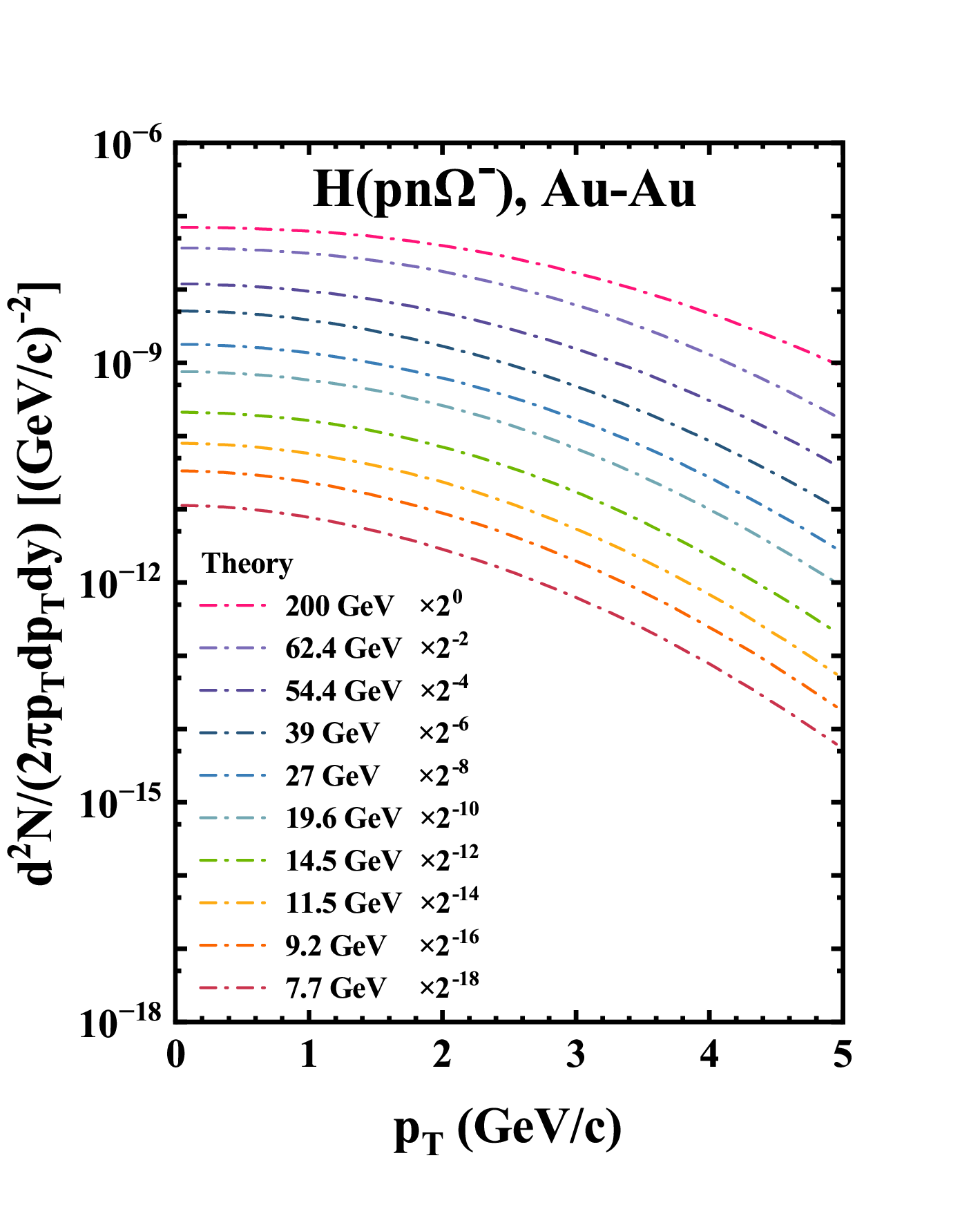}\\
	\caption{Predictions of invariant $p_T$ spectra of $H(pn\Omega^-)$ in $0-10\%$ centrality of Au-Au collisions at $\sqrt{s_{NN}}$ = 7.7, 9.2, 11.5, 14.5, 19.6, 27, 39, 54.4, 62.4 and 200 GeV. }
	\label{fig:pnO-pt}
\end{figure}

Figures~\ref{fig:pOnO-pt} and \ref{fig:pnO-pt} show predictions for $p_T$ spectra of $H(p\Omega^-)$, $H(n\Omega^-)$ and $H(pn\Omega^-)$ in $0-10\%$ centrality of Au-Au collisions at $\sqrt{s_{NN}}=7.7$, 9.2, 11.5, 14.5, 19.6, 27, 39, 54.4, 62.4 and 200 GeV.
Different lines are scaled by different factors for clarity at different collision energies as shown in the figures.

\begin{table*}
\centering
\caption{Predictions of yield densities $dN/dy$ and average transverse momentum $\langle{p_T}\rangle$ of $H(p\Omega^-)$, $H(n\Omega^-)$, and $H(pn\Omega^-)$ in Au-Au collisions in $0-10\%$ centrality at $\sqrt{s_{NN}}=7.7$, 9.2, 11.5, 14.5, 19.6, 27, 39, 54.4, 62.4 and 200 GeV.}
\renewcommand{\arraystretch}{1}
\setlength{\tabcolsep}{10pt}

\begin{tabular}{ccccccccc}
\toprule
\multirow{2}{*}{$\sqrt{s_{NN}}$ (GeV)} &\multicolumn{2}{c}{$H(p\Omega^-)$} && \multicolumn{2}{c}{$H(n\Omega^-)$}&& \multicolumn{2}{c}{$H(pn\Omega^-)$}\\
\cline{2-3}\cline{5-6}\cline{8-9}
&$dN/dy$ (×$10^{-4}$) & $\langle{p_T}\rangle$ (GeV/c) && $dN/dy$ (×$10^{-4}$) & $\langle{p_T}\rangle$ (GeV/c) && $dN/dy$ (×$10^{-5}$) & $\langle{p_T}\rangle $\\
\hline
$7.7$ & $7.829$&$1.281$&&$7.402$&$1.275$&&$2.735$&$1.569$\\
$9.2$ & $7.496$&$1.295$&&$7.061$&$1.289$&&$2.091$&$1.586$\\
$11.5$ &$6.641$&$1.323$&&$6.236$&$1.318$&&$1.342$&$1.623$\\
$14.5$ &$6.465$&$1.368$&&$5.992$&$1.362$&&$0.984$&$1.671$\\
$19.6$ & $7.942$&$1.392$&&$7.340$&$1.386$&&$0.909$&$1.704$\\
$27$ &$6.306$&$1.414$&&$5.754$&$1.408$&&$0.542$&$1.729$\\
$39$ &$6.046$&$1.425$&&$5.498$&$1.419$&&$0.380$&$1.738$\\
$54.4$ & $5.152$&$1.512$&&$4.660$&$1.505$&&$0.263$&$1.843$\\
$62.4$ &$4.945$&$1.574$&&$4.435$&$1.567$&&$0.235$&$1.926$\\
$200$ &$4.545$&$1.725$&&$4.087$&$1.716$&&$0.139$&$2.102$\\
\hline\hline
\label{table:dNdypt-Ohyp}
\end{tabular}
\end{table*}

Table \ref{table:dNdypt-Ohyp} presents predictions of the average transverse momenta $\langle p_T\rangle$ and yield rapidity densities $dN/dy$ of $H(p\Omega^-)$, $H(n\Omega^-)$, and $H(pn\Omega^-)$ in Au-Au collisions in $0-10\%$ centrality at $\sqrt{s_{NN}}=7.7$, 9.2, 11.5, 14.5, 19.6, 27, 39, 54.4, 62.4 and 200 GeV.
For $\langle p_T\rangle$, there is an obviously increasing behavior from $\sqrt{s_{NN}}=7.7$ to 200 GeV, similar to light nuclei and $^3_{\Lambda}$H.
For $dN/dy$, both $H(p\Omega^-)$ and $H(n\Omega^-)$ exhibit decreasing trend as a function of collision energy, while $H(pn\Omega^-)$ decreases more seriously.
This is because the decreasing nucleons with the increasing $\sqrt{s_{NN}}$ doubly affect $H(pn\Omega^-)$ production compared to $H(p\Omega^-)$ and $H(n\Omega^-)$.
The slightly lower results of $H(n\Omega^-)$ than $H(p\Omega^-)$ come from its slightly larger size.
Our predicted productivities of $H(p\Omega^-)$, $H(n\Omega^-)$, and $H(pn\Omega^-)$ at $\sqrt{s_{NN}}=200$ GeV are in the same magnitude as those in Refs.~\cite{Zhang:2020dma,Zhang:2021vsf}.
Predictions in other collision energies in Table \ref{table:dNdypt-Ohyp} provide more detailed references for future search and measurement of these $\Omega-$hypernuclei in different experiments.

\section{Production correlations of light (hyper-)nuclei} \label{Cor-results}

Yield ratios and average transverse momentum relations are two groups of global correlations, which carry intrinsic production relationships among different light (hyper-)nuclei.
Such correlations are regarded as characteristic probes of production mechanisms of composite nuclei~\cite{Sun:2018mqq,Wang:2023rpd}.
Based on theoretical results in Sec.~\ref{pTresults}, we study production correlations of various species of nuclei in both light and strange sectors in this section.

\subsection{Yield ratios of light (hyper-)nuclei}      

\begin{figure*}[htbp]
	\centering
	\includegraphics[width=1\linewidth]{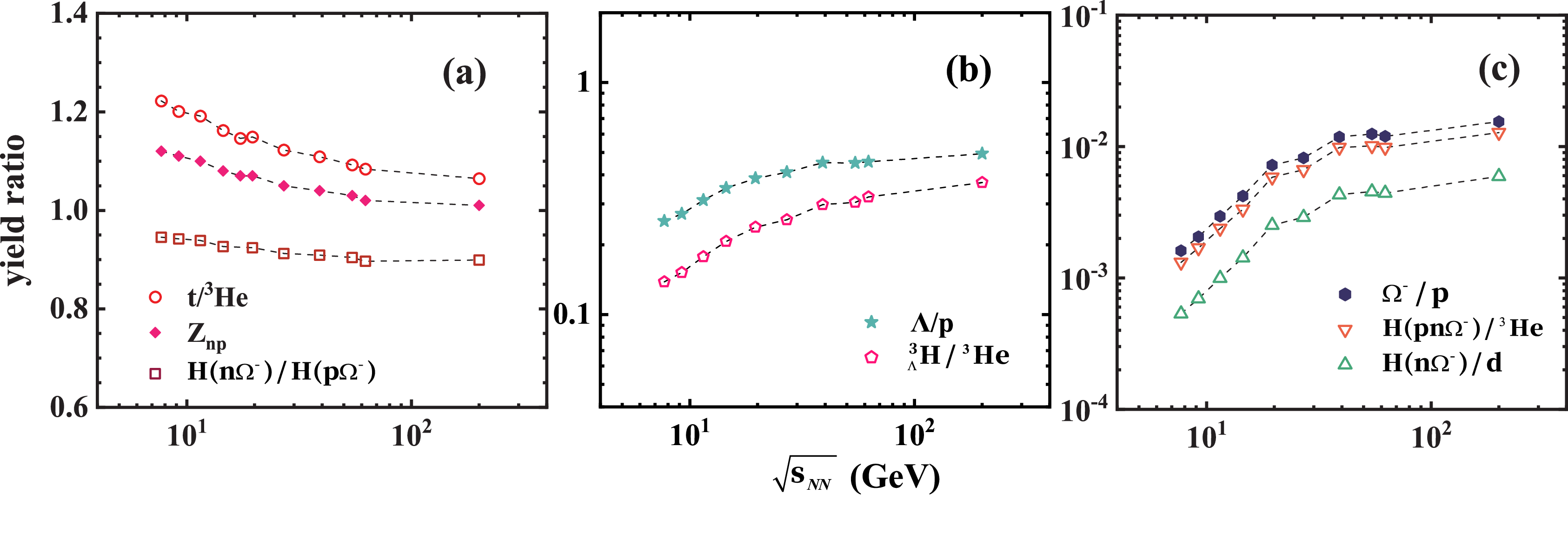}\\
	\caption{Yield ratios (a) $t/^3\text{He}$, $H(n\Omega^-)/H(p\Omega^-$), $Z_{np}$, (b) $_\Lambda^3\text{H}/^3\text{He}$, $\Lambda/p$, and (c) $H(pn\Omega^-)/^3\text{He}$, $H(n\Omega^-)/d$, $\Omega^-/p$ as a function of $\sqrt{s_{\text{NN}}}$ in central Au-Au collisions. 
Different symbols connected by dashed lines to guide the eye are theoretical results.}
	\label{fig:ratio1}
\end{figure*}

We first study yield ratios $t/^3\text{He}$ and $H(n\Omega^-)/H(p\Omega^-)$, both of which possess the same constitute ratio $n/p$. 
In the coalescence framework, they should have similar behaviors as the primordial nucleon yield ratio $n/p$, i.e., $Z_{np}$ defined in Sec.~\ref{np-pTdistribution}.
Fig.~\ref{fig:ratio1} (a) shows the results of $t/^3\text{He}$ and $H(n\Omega^-)/H(p\Omega^-)$ as well as $Z_{np}$ as a function of collision energy $\sqrt{s_{NN}}$.
Open circles and squares are for $t/^3\text{He}$ and $H(n\Omega^-)/H(p\Omega^-)$, respectively.
Filled rhombuses are for $Z_{np}$.
They are all connected by dashed lines to guide the eye.
Both $t/^3\text{He}$ and $H(n\Omega^-)/H(p\Omega^-)$ decrease with the increasing $\sqrt{s_{NN}}$, exhibiting the same trend as $Z_{np}$, but they separate on different sides of $Z_{np}$.
The up side location of $t/^3\text{He}$ results from the smaller size of $t$ than that of $^3\text{He}$, which leads to weaker production suppression for $t$ compared to $^3\text{He}$.
For $H(n\Omega^-)/H(p\Omega^-)$, the numerator $H(n\Omega^-)$ has larger size, and its production suppression is stronger than $H(p\Omega^-)$.
This makes $H(n\Omega^-)/H(p\Omega^-)$ locate at the down side of $Z_{np}$.

Fig.~\ref{fig:ratio1} (b) presents $^3_{\Lambda}\text{H}/^3\text{He}$ and $\Lambda/p$, which are denoted by open pentagons and filled stars.
Fig.~\ref{fig:ratio1} (c) gives $H(pn\Omega^-)/^3\text{He}$ and $H(n\Omega^-)/d$ as well as $\Omega^-/p$, which are denoted by open triangles as well as filled hexagons.
The increasing yields of hyperons and the decreasing yields of nucleons make $\Lambda/p$ and $\Omega^-/p$ increase as a function of $\sqrt{s_{NN}}$.
This dominates the increasing trend for $^3_{\Lambda}\text{H}/^3\text{He}$, $H(pn\Omega^-)/^3\text{He}$, and $H(n\Omega^-)/d$.
The down side of nucleus ratio compared to yield ratios of hyperons to protons comes from the larger sizes of the nuclei in the numerator than those in the denominator.
All the results in Fig.~\ref{fig:ratio1} tell one that yield ratios of different nuclei exhibit the same behaviors as the corresponding primordial baryon yield ratios.
This is the natural result in the hadronic coalescence picture.
A more interesting fact is that the locations of nucleus ratios are determined by the relative size of the nuclei in numerator and denominator.
The smaller size of the numerator nucleus gives the nucleus ratio upside while the larger size gives downside.
This provides a powerful method to discriminate the relative sizes of different nuclei via their productivities.

\begin{figure}[htbp]
	\centering
	\includegraphics[width=0.98\linewidth]{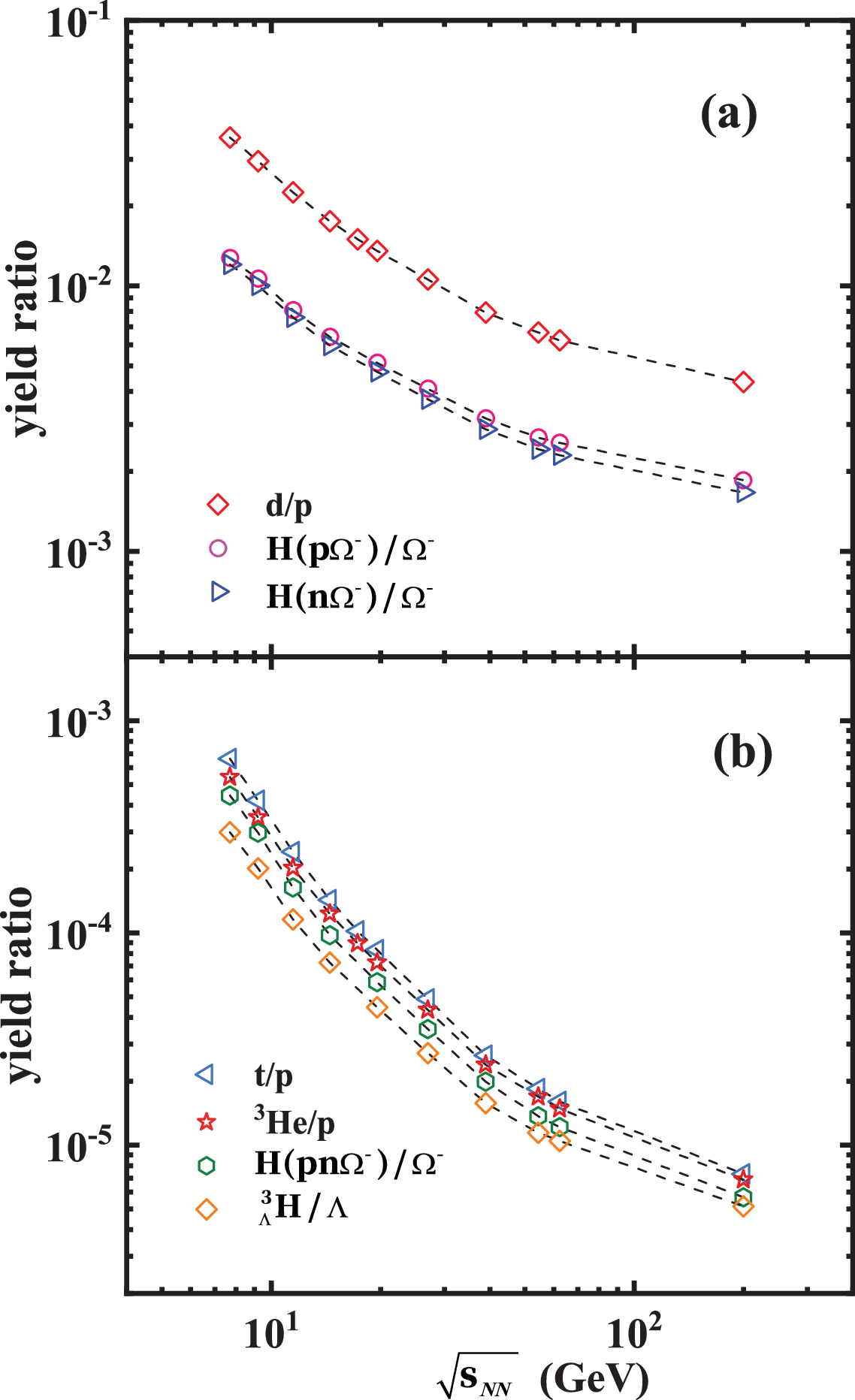}\\
	\caption{Yield ratios (a) $d/p$, $H(p\Omega^-)/\Omega^-$, $H(n\Omega^-)/\Omega^-$, and (b) $t/p$, $^3\text{He}/p$, $H(pn\Omega^-)/\Omega^-$, $_{\Lambda}^3\text{H}/\Lambda$ as a function of $\sqrt{s_{\text{NN}}}$ in central Au-Au collisions. 
Different symbols connected by dashed lines to guide the eye are theoretical results.}
	\label{fig:ratio2}
\end{figure}

Figure~\ref{fig:ratio2} shows the yield ratios of different nuclei to protons or hyperons.
Open rhombuses, circles, and triangles connected by dashed lines to guide the eye in Fig.~\ref{fig:ratio2} (a) are the theoretical results of $d/p$, $H(p\Omega^-)/\Omega^-$, and $H(n\Omega^-)/\Omega^-$, respectively.
Open triangles, stars, rhombuses, and hexagons connected by dashed lines to guide the eye in Fig.~\ref{fig:ratio2} (b) are the theoretical results of $t/p$, $^3\text{He}/p$, $^3_{\Lambda}\text{H}/\Lambda$, and $H(pn\Omega^-)/\Omega^-$, respectively.
These ratios tend to offset the differences from the primordial nucleons and hyperons, making them potent to reveal the underlying universality of the production of different species of nuclei in light and
strange sectors.
The vestigial constitutes of the above ratios are just nucleons.
So all of these ratios decrease with the increasing collision energy.
The severely decreasing trend in Fig.~\ref{fig:ratio2} (b) compared to that in Fig.~\ref{fig:ratio2} (a) is due to the remanent double nucleon constitutes in the yield ratios of nuclei with atomic mass number $A=3$ to nucleons or hyperons.

\subsection{Average transverse momentum relations}

\begin{figure}[htbp]
	\centering
	\includegraphics[width=0.45\textwidth]{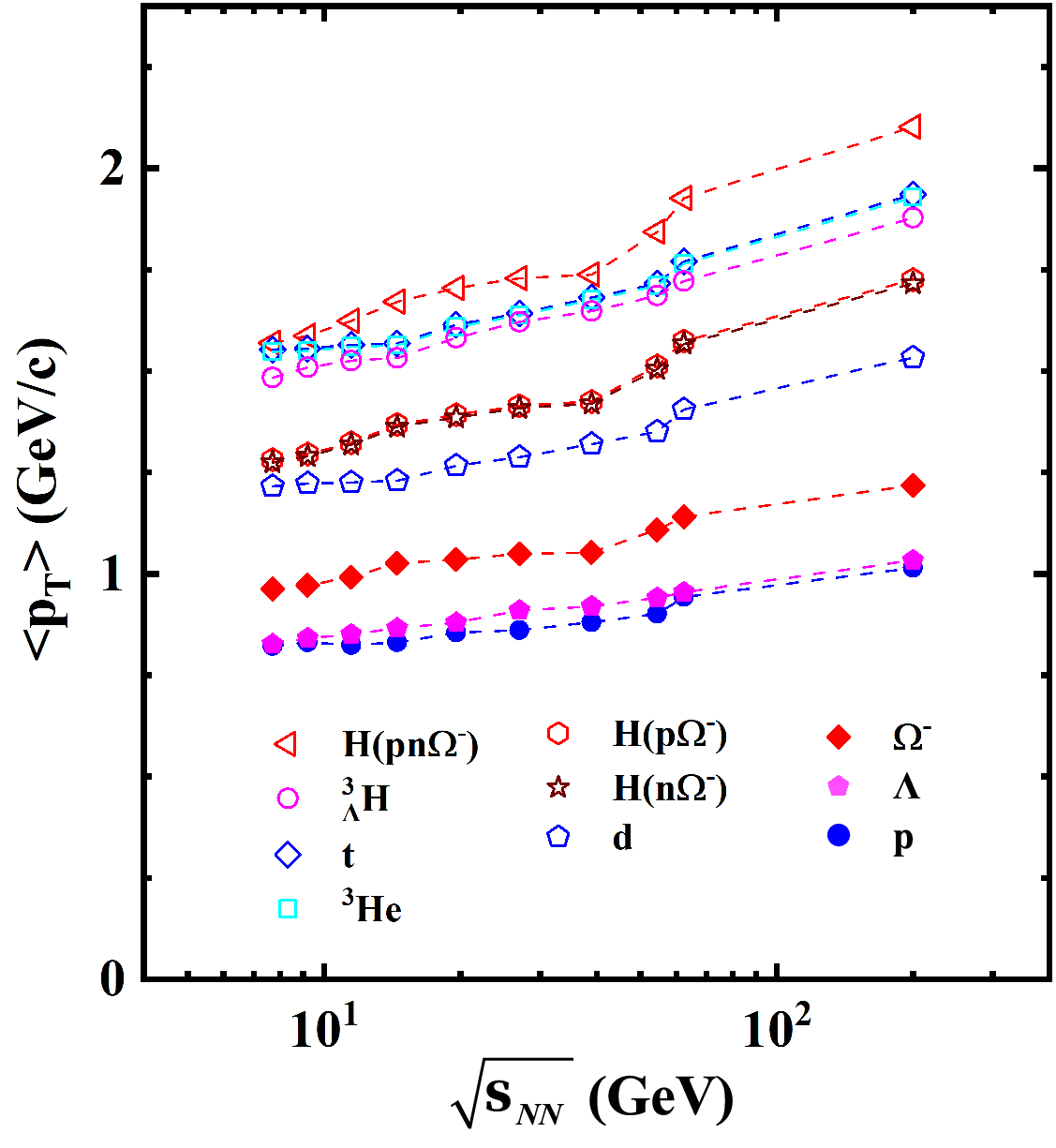}
	\caption{$\langle p_T \rangle$ of light (hyper-)nuclei as well as baryons $p$, $\Lambda$, $\Omega^-$ as a function of $\sqrt{s_{NN}}$.}
	\label{fig:meanpT}
\end{figure}

The average transverse momentum $\langle p_T \rangle$ of a specific species of particles is mainly determined by the $p_T$ spectrum in the low and intermediate $p_T$ range, and so it can reflect the collective property of soft particles.
Here we study the $\langle p_T \rangle$ relations of different species of light (hyper-)nuclei as well as ordinary baryons $p$, $\Lambda$, $\Omega^-$.

Figure~\ref{fig:meanpT} presents our theoretical results of $\langle p_T \rangle$ of different light (hyper-)nuclei as well as baryons $p$, $\Lambda$, $\Omega^-$ as a function of $\sqrt{s_{NN}}$.
Filled circles, pentagons, and rhombuses are for $p$, $\Lambda$, and $\Omega^-$, respectively.
They exhibit a distinct mass order, i.e., $\langle p_T \rangle_p < \langle p_T \rangle_{\Lambda} < \langle p_T \rangle_{\Omega^-}$.
This mass-ordering behavior is usually regarded as a signal of hadronic collective motion and can be naturally explained by hydrodynamical models such as in Ref.~\cite{Kolb:2002ve}.
Open pentagons, hexagons, and stars are for $d$, $H(p\Omega^-)$, and $H(n\Omega^-)$, respectively.
They also obey the mass order $\langle p_T \rangle_d < \langle p_T \rangle_{H(p\Omega^-)} \sim \langle p_T \rangle_{H(n\Omega^-)}$.
The negligible mass difference and very small size difference of $H(p\Omega^-)$ and $H(n\Omega^-)$ give nearly the same average transverse momenta.
For nuclei with $A=3$, the results of $^3\text{He}$, $t$, $^3_\Lambda\text{H}$, and $H(pn\Omega^-)$ are denoted by open squares, rhombuses, circles, and triangles, respectively, which give $\langle p_T \rangle_{^3_\Lambda\text{H}} < \langle p_T \rangle_{^3\text{He}} \sim \langle p_T \rangle_t < \langle p_T \rangle_{H(pn\Omega^-)}$.
The mass order is violated by $^3_\Lambda\text{H}$.
This violation comes from the relatively large size of $^3_\Lambda\text{H}$, accompanying with the softening of its $p_T$ spectrum.
Such softening effect from the large $^3_\Lambda\text{H}$ size was firstly proposed in Ref.~\cite{Liu:2024ygk}.
It can be clearly decoded by the Lorentz $\gamma$ factor in Eqs.~(\ref{eq:pt-Hj2h}) and (\ref{eq:pt-Hj3h}).
The $\gamma$ hardens the nucleus $p_T$ distribution, which is firstly proposed to understand the increasing behaviors of the coalescence factor $B_A$ as a function of $p_T$~\cite{Wang:2020zaw}.
But this hardening effect is suppressed by the size of nucleus itself.
The larger size gives stronger suppression, and then soften the $p_T$ spectrum.
This softening effect also exists in $\Omega-$hypernuclei, but it is not strong enough to violate the mass order.

\section{summary}    \label{summary}

We considered the isospin asymmetry to extend the analytical coalescence model previously developed at the LHC for the production of light (hyper-)nuclei at midrapidity.
The formulae of momentum distributions of light (hyper-)nuclei in two-body coalescence and three-body coalescence were derived. 
The relationships of light (hyper-)nuclei with primordial nucleons and hyperons in momentum space in the laboratory frame were presented.
The influences of the hadronic system scale and the nucleus own size on the nucleus production were clearly given.

We applied the extended coalescence model to Au-Au collisions at the RHIC, based on nucleons and hyperons created at kinetic freeze-out from the QCM, to study production properties of nuclei in light and strange sectors.
The available experimental data of $p_T$ spectra of $d$, $t$, $^3$He, and $^3_{\Lambda}$H measured by the STAR collaboration in Au-Au collisions at $\sqrt{s_{NN}}=7.7,~9.2,~11.5,~14.5,~17.3,~19.6,~27,~39,~54.4,~62.4,~200$ GeV were explained.
The predictions of the $p_T$ spectra of $H(p\Omega^-)$, $H(n\Omega^-)$ and $H(pn\Omega^-)$ were also provided for future experimental measurements.

Notably, we studied production correlations of different species of light (hyper-)nuclei.
We discussed two groups of correlations.
One referred to yield ratios.
We compared the behaviors of yield ratios of different nuclei as a function of collision energy with the corresponding yield ratios of primordial nucleons and hyperons.
We found the locations of nucleus yield ratios compared to nucleon/hyperon yield ratios were uniquely determined by the relative size of the nuclei.
This provided a new way to discriminate the relative sizes of different nuclei via their relative productivities.
The other group referred to average transverse momentum relations.
We found the $\langle p_T \rangle$'s of all the light (hyper-)nuclei except the $^3_{\Lambda}$H obey the mass order.
The mass order violation of $^3_{\Lambda}$H resulted from its own large size as well as relatively small $\langle p_T \rangle$ difference of nucleons and $\Lambda$ hyperons.
All these results set new light to production characteristics of light nuclei and hyper-nuclei in high energy heavy-ion collisions.

\section*{Acknowledgements}

We thank profs. Xiao-Feng Luo, Jun Song and Hui Liu for helpful discussions.
We also thank the STAR collaboration for providing us with the experimental data of $d$ and $^3$He in BES II.
This work was supported in part by the National Natural Science Foundation of China under Grants No. 12375074 and No. 12175115, and the Natural Science Foundation of Shandong
Province, China, under Grant No. ZR2025QC38.

\bibliographystyle{apsrev4-1}
\bibliography{myref}

\end{document}